\documentclass[letterpaper]{JHEP3} 
\pdfoutput=1
\usepackage{amsmath,amssymb,amsfonts,amsthm,dsfont,dcolumn,yfonts}
\usepackage{tensor, slashed}

\usepackage{epsfig,multicol}

\newcommand\fverb{\setbox\fverbbox=\hbox\bgroup\verb}
\newcommand\fverbdo{\egroup\medskip\noindent%
			\fbox{\unhbox\fverbbox}\ }
\newcommand\fverbit{\egroup\item[\fbox{\unhbox\fverbbox}]}
\newbox\fverbbox

\newcommand{\be}{\begin{equation}}
\newcommand{\ee}{\end{equation}}
\newcommand{\bea}{\begin{eqnarray}}
\newcommand{\eea}{\end{eqnarray}}
\newcommand{\beq}{\begin{equation}}
\newcommand{\eeq}{\end{equation}}
\newcommand{\beqn}{\begin{eqnarray}}
\newcommand{\eeqn}{\end{eqnarray}}
\def\ie{{\it i.e.~}}

\def\det{{\rm{det}}}

\def\ads{{\rm AdS}}

\def\G{\Gamma}

\newcommand{\wn}{\textswab{w}\kern1pt}
\newcommand{\qn}{\textswab{q}\kern1pt}

\title{Holography and the sound of criticality}

\author{Mohammad Edalati, Juan I. Jottar, Robert G. Leigh \\ Department of Physics, University of Illinois, 1110 W. Green Street, Urbana, IL 61801, U.S.A.\\ \\E-mail: \email{edalati@illinois.edu}, \email{jjottar2@illinois.edu}, \email{rgleigh@illinois.edu}}

\abstract{Using gauge/gravity duality techniques, we discuss the sound-channel retarded correlators of vector and tensor conserved currents in a class of $(2+1)$-dimensional strongly-coupled field theories at zero temperature and finite charge density, assumed to be holographically dual to the extremal Reissner-Nordstr\"{o}m AdS$_4$ black hole. Using a combination of analytical and numerical methods, we determine the quasinormal mode spectrum at finite momentum for the coupled gravitational and electromagnetic perturbations, and discuss the appropriate choice of gauge-invariant variables (master fields) in order for the black hole quasinormal frequencies to reproduce the field theory spectrum. We discuss the role of the near horizon AdS$_{2}$ geometry in determining the low-frequency behavior of retarded correlators in the boundary theory, and comment on the emergence of criticality in the IR. In addition, we establish the existence of a sound mode at zero temperature and compute the speed of sound and sound attenuation constant numerically, obtaining a result consistent with the expectations from the zero temperature limit of hydrodynamics. The dispersion relation of higher resonances is also investigated.}


\begin{document} 
%%%%%%%%%%%%%%%%%%%%%%%%%%%%%%%%%%%%%%%%%%%%%%%%%%%%%%%%%%%%%%%%%%%%%%%%%%%%%%%%%%%%%%%%%%
%%%%%%%%%%%%%%%%%%%%%%%%%%%%%%%%%%%%%%%%%%%%%%%%%%%%%%%%%%%%%%%%%%%%%%%%%%%%%%%%%%%%%%%%%%
\section{Introduction}
The recent surge in applications of the AdS/CFT correspondence in modeling strongly-interacting theories in the same universality class of systems relevant to condensed matter physics is an indication of gauge/gravity duality techniques being a promising new approach to study these systems.\footnote{See \cite{Hartnoll:2009sz, Herzog:2009xv, McGreevy:2009xe, Horowitz:2010gk} for accounts of  applications in condensed matter physics.} 
One particular such avenue, which we focus on in this paper, uses phenomenological classical gravity plus matter systems in order to study some aspects of the quantum critical regions in the phase diagram of condensed matter systems via holography. At a quantum critical point, the system undergoes a quantum phase transition at zero temperature and it is often modelled by a strongly-coupled conformal field theory, a regime where a standard field theoretic understanding is generically lacking. One may then hope that the AdS/CFT correspondence will be useful in addressing questions that go beyond the scope of the usual perturbative techniques. 

In this paper, we focus on a class of strongly-coupled $(2+1)$-dimensional field theories at zero temperature and finite U(1) charge density. The simplest gravity background that captures the essential features of such theories in a holographic setup is the extremal Reissner-Nordstr\"{o}m AdS$_{4}$ (RN-AdS$_{4}$) black hole. By studying the coupled electromagnetic and gravitational fluctuations about the RN-AdS$_{4}$ black hole, the AdS/CFT correspondence allows us to calculate the retarded correlators of the U(1) charge current ($J^{\mu}$) and energy-momentum ($T^{\mu}_{\phantom{\mu}\nu}$) operators of the dual field theory for non-vanishing frequency and momentum. This holographic system has recently attracted much attention in the literature, inasmuch as the dual field theory exhibits a variety of emergent quantum critical phenomena, including non-Fermi liquid type behavior in the case of fermionic operators \cite{Faulkner:2009wj}. The emergence of the critical features in the IR physics of the dual theory is rooted in the fact that the near horizon region of the extremal background geometry is ${\rm AdS}_2\times\mathds{R}^2$, and the assumption that there exists a so-called IR CFT dual to the AdS$_2$ factor. 

As a complement to our previous studies \cite{Edalati:2009bi,Edalati:2010hk} of transport coefficients and transverse (shear) channel correlators of $J^{\mu}$ and $T^{\mu}_{\phantom{\mu}\nu}$ at finite frequency and momentum, here we explore the structure of correlators of the vector and tensor currents in the longitudinal (sound) channel of the dual field theory. In particular, by using a combination of analytical and numerical techniques, we determine the spectrum of excitations for generic values of the (spatial) momentum. Along the way, we comment on the appropriate choice of gauge-invariant bulk variables (master fields), such that their associated quasinormal frequencies encode the spectrum of excitations of the dual theory. Studying the retarded correlators at finite frequency and momentum, we present evidence for the existence of a branch cut on the negative imaginary axis, as well as a series of metastable modes corresponding to isolated poles in the lower-half complex frequency plane. At low frequencies (as compared to the chemical potential $\mu$ conjugate to the charge density), but generic momentum, we find that the retarded correlators of the sound-channel vector and tensor operators exhibit emergent scaling behavior. As in the case of the shear-channel modes studied in \cite{Edalati:2009bi,Edalati:2010hk}, these operators give rise in the IR to two sets of scalar operators with different conformal dimensions. 

Turning to the low frequency and momentum regime, we establish the existence of collective sound modes at zero temperature, and compute the associated speed of sound and sound attenuation constant. We compare our results with the expectations from the zero-temperature limit of well-known hydrodynamical relations, and discuss the extent to which both results agree. Similarly, we study the dispersion relations of higher resonances, which are fundamentally different from that of the sound mode: their frequencies $\omega$ remain finite as the momentum $k$ approaches zero, so that the spectrum for $\omega \ll \mu$ and $k \ll \mu$ is effectively dominated by the sound mode. We provide further evidence to support this picture by computing the sound poles residue and comparing the Green's functions obtained numerically with a truncated spectral representation including the sound modes only, observing an excellent agreement for values of the momenta in the range $0 \leq k \leq \mu$. 

This paper is organized as follows. In section \ref{section: field theory} we review some notions concerning linear response theory (LRT), hydrodynamics, and the sound modes, and comment on the applicability of these concepts to the zero-temperature scenario which is the main focus of this work. In section \ref{section: background} we review the RN-AdS$_{4}$ black hole background, its near horizon ${\rm AdS}_2\times\mathds{R}^2$ geometry, and the basic properties of the dual field theory. We then present the linearized Einstein-Maxwell system and introduce the gauge-invariant combinations known as ``master fields", which allow us to reduce the system to only two degrees of freedom satisfying decoupled second order differential equations. In section \ref{section: Green's functions and criticality} we exhibit the asymptotic solution for the various bulk modes, and redefine the master fields in order to obtain a new set of gauge-invariant variables which is better suited to the calculation of the holographic Green's functions. We then present formal expressions for the complete matrix of retarded correlators in the longitudinal channel, and elucidate their low-energy behavior by using the method of matching of asymptotic expansions, finding emergent scaling behavior of the spectral functions for generic momentum. Using the decoupled equations introduced in section \ref{section: Green's functions and criticality}, in section \ref{section: spectrum} we numerically compute the quasinormal modes of the extremal RN-AdS$_{4}$  black hole, which by virtue of our choice of master fields coincide with the dual field theory spectrum at zero temperature. Turning to the small frequency and momentum regime, we establish the existence of sound modes at zero temperature, and compute the speed of sound and sound attenuation constant numerically. Similarly, we investigate the dispersion relation of higher resonances (overtones) and discuss the extent to which the sound poles dominate the spectrum in the IR by computing their residues. We conclude in section \ref{section: conclusions}. 
%%%%%%%%%%%%%%%%%%%%%%%%%%%%%%%%%%%%%%%%%%%%%%%%%%%%%%%%%%%%%%%%%%%%%%%%%%%%%%%%%%%%%%%%%%
%%%%%%%%%%%%%%%%%%%%%%%%%%%%%%%%%%%%%%%%%%%%%%%%%%%%%%%%%%%%%%%%%%%%%%%%%%%%%%%%%%%%%%%%%%
\section{Field theory generalities}\label{section: field theory}
Even though we will study the field theory spectrum at generic, not necessarily small\footnote{As compared to the chemical potential $\mu$.}, values of the frequency and momenta, in order to interpret our results it is instructive to review some notions concerning the hydrodynamic (sound) modes, and comment on the applicability of these concepts to the zero temperature case which concerns us here.
%%%%%%%%%%%%%%%%%%%%%%%%%%%%%%%%%%%%%%%%%%%%%%%%%%%%%%%%%%%%%%%%%%%%%%%%%%%%%%%%%%%%%%%%%%
\subsection{Finite temperature: LRT and the sound modes}
Here we will discuss only a few basic facts that will be relevant to our exposition. Detailed discussions of hydrodynamics in the context of the AdS/CFT correspondence can be found in \cite{Son:2007vk,Amado:2008ji}, for example. From the field theory point of view, the hydrodynamical regime is an effective description valid at wavelengths which are much larger than the mean free path of individual particles in the system. Among the various excitations in the spectrum, a prominent role is played by the so-called hydrodynamic modes, whose dispersion relation satisfies $\omega_{\mbox{\tiny{hydro}}}(k \to 0) \to 0$. The importance of these modes becomes apparent in linear response theory (LRT), where the response of a field $\phi(t,\vec{r})$ (assumed to be a scalar field for simplicity) to a perturbation by an external source $j(t,\vec{r})$ is given by 
\begin{equation}
\langle \phi(t,\vec{r})\rangle = -\int dt' d^{2}\vec{r}^{\, '}\,G_{R}(t - t', \vec{r} -\vec{r}^{\, '})j(t',\vec{r}^{\, '})\, ,
\end{equation}

\noindent where $G_{R}(t - t', \vec{r} - \vec{r}^{\, '})$ is the retarded two-point function given by
\begin{equation}
G_{R}(t - t', \vec{r} - \vec{r}^{\, '}) = -i\theta(t-t')\left\langle[\phi(t,\vec{r}),\phi(t',\vec{r}^{\, '})]\right\rangle.
\end{equation}

\noindent We will denote by $\{\omega_{\star}\}$ the set of poles of $G_{R}(t - t', \vec{r} - \vec{r}^{\, '})$ in the complex frequency plane, which define the spectrum of excitations of the theory. For simplicity, we assume that these poles are simple, and that there are no additional non-analyticities. Going to momentum space, we can then represent the retarded two-point function as
\begin{equation}\label{spectral decomposition}
G_{R}(\omega,\vec{k}) = \sum_{\omega_{j} \in \{\omega_{\star}\}}\frac{\mbox{Res}(\omega_{j},\vec{k})}{\omega - \omega_{j}(\vec{k})} + \mbox{terms analytic in }(\omega,\vec{k})\, ,
\end{equation} 

\noindent and the response of the field to the perturbation represented by $j(t',\vec{r}^{\, '})$ takes the form 
\begin{equation}
\langle \phi(t,\vec{r})\rangle = i\theta(t)\int \frac{d^{2}\vec{k}}{(2\pi)^{2}}e^{i\vec{k}\cdot \vec{r}}\sum_{\omega_{j} \in \{\omega_{\star}\}}e^{\mbox{\footnotesize{Im}}(\omega_{j})\, t - i\,\mbox{\footnotesize{Re}}(\omega_{j})\, t}\,\mbox{Res}(\omega_{j},\vec{k})\, j(\omega_{j},\vec{k})\, .
\end{equation}
 
\noindent Thus, $\mbox{Re}(\omega_{j})$ and $\mbox{Im}(\omega_{j})$ determine the frequency and damping, respectively, of the mode associated with the eigenfrequency $\omega_{j}$. In particular, eigenfrequencies with $\mbox{Im}(\omega_{j}) >0$ produce instabilities that signal the breakdown of the linear response approximation. The eigenmodes with $\mbox{Im}(\omega_{j}) <0$ represent the dissipative response of the system to the external perturbation and they are referred to as \textit{quasinormal modes} (QNMs). Similarly, the residue $\mbox{Res}(\omega_{j},\vec{k})$ determines the weight of the corresponding contribution to the total response. The important observation is that for small values of the momenta the hydrodynamic modes have, by definition, the smallest imaginary part and therefore they dominate the relaxation of the system. In fact, for sufficiently large time scales\footnote{See \cite{Amado:2008ji} for a discussion of the relevant time scale and the importance of the quasinormal mode residues in determining it.} the total response can be accurately approximated by the contribution of the hydrodynamic modes only. In particular, in the longitudinal (sound) channel one finds two hydrodynamic modes. Assuming rotational invariance in the spatial $\vec{r}=(x,y)$ plane, we can align the momentum with the $x$-axis, i.e.  $\vec{k}=k\,\hat{x}$; if the underlying theory has unbroken parity ($y \to -y$) invariance, these two poles have the same imaginary part and opposite sign real parts, and obey a dispersion relation of the form
\begin{equation}\label{sound modes dispersion}
\omega_{s}(k) = c_{s}k -i \Gamma_{s} k^{2} + \mathcal{O}(k^{3}),
\end{equation}
at small momentum, where $c_{s}$ is the \textit{speed of sound} and $\Gamma_{s}$ is the \textit{sound attenuation} constant. Denoting the residue associated with $\omega_{s}$ by $\mbox{R}_{s}(k)$, the representation \eqref{spectral decomposition} for the correlator in the hydrodynamic regime can then be approximated by
\begin{equation}\label{Breit Wigner in sound channel}
G^{\mbox{\tiny{sound}}}_{R}(\omega, k \to 0) \simeq \frac{\mbox{R}_{s}(k)}{\omega - \omega_{s}(k)} +\frac{-\mbox{R}^{*}_{s}(k)}{\omega + \omega^{*}_{s}(k)} + \mbox{terms analytic in }(\omega , k)\, .
\end{equation} 

\noindent As one increases the momentum the effects of higher resonances become important, and the expression above should be modified accordingly.

%%%%%%%%%%%%%%%%%%%%%%%%%%%%%%%%%%%%%%%%%%%%%%%%%%%%%%%%%%%%%%%%%%%%%%%%%%%%%%%%%%%%%%%%%%
\subsection{Zero temperature}\label{zero temperature hydro}
While it is tempting to use the familiar language introduced above, one should keep in mind that the main focus in this paper is the zero temperature scenario, where it is not clear in what precise sense one can talk about hydrodynamics. In this light, our results should be primarily understood as an exploration of the structure of retarded correlators of conserved currents in the longitudinal channel of a $(2+1)$-dimensional strongly-coupled field theory at zero temperature and finite charge density. As discussed in \cite{Faulkner:2009wj,Edalati:2009bi,Edalati:2010hk}, additional care has to be exercised when studying the zero temperature dual of the extremal RN-AdS$_{4}$ black hole. This is due to the fact that the $T\to 0$ and $\omega \to 0$ limits do not, in general, commute. This implies that the naive $T\to 0$ limit of some finite temperature expressions does not agree with the corresponding quantities computed by setting $T=0$ initially. Additionally, the analytic structure of retarded correlators becomes richer; for example, isolated poles at zero temperature can now coalesce into branch cuts. 

While being cognizant of the subtleties mentioned above, in later sections we will find that some of our results do in fact agree with the $T \to 0$ limit of certain hydrodynamic equations. In particular, at $T =0$ the notion of ``hydrodynamic limit" entails considering frequencies and momenta which are small compared to the scale set by the chemical potential $\mu$. In this regime, we establish the existence of exactly two modes with a dispersion relation of the form \eqref{sound modes dispersion}, with the constants $c_{s}$ and $\Gamma_{s}$ matching, within the numeric precision, those predicted by taking the $T \to 0$ limit of well-known hydrodynamic expressions. This justifies referring to these excitations in the zero-temperature theory as ``sound modes".  Just like in the finite temperature case, we will see that these modes effectively dominate the spectrum at small frequency and momenta: the retarded Green's functions computed numerically via holographic techniques will be shown to be in excellent agreement with the approximation \eqref{Breit Wigner in sound channel}. In fact, the fitting is robust not only for $k \ll \mu$, but also for values of the momenta of the order of the chemical potential. For larger values of the momenta it becomes necessary to include the effect of higher resonances in order for the truncated spectral representation of the correlator to be accurate. 
%%%%%%%%%%%%%%%%%%%%%%%%%%%%%%%%%%%%%%%%%%%%%%%%%%%%%%%%%%%%%%%%%%%%%%%%%%%%%%%%%%%%%%%%%%
%%%%%%%%%%%%%%%%%%%%%%%%%%%%%%%%%%%%%%%%%%%%%%%%%%%%%%%%%%%%%%%%%%%%%%%%%%%%%%%%%%%%%%%%%%
\section{The gravity background and the linearized theory}\label{section: background}
Consider the Einstein-Maxwell action in $3+1$ spacetime dimensions with a negative cosmological constant $\Lambda = -3/L^2$
\bea\label{EMaction}
S= \frac{1}{2\kappa_4^2}\int d^4x \sqrt{-g}\left(R-2\Lambda- L^2F_{\mu\nu}F^{\mu\nu}\right)\, ,
\eea

\noindent where $L$ is the curvature radius of AdS$_4$. The background we are interested in is the Reissner-Nordstr\"{o}m AdS$_4$ (RN-AdS$_{4}$) black hole
\begin{align}\label{metric}
ds^2&=g_{\mu\nu}dx^\mu dx^\nu=\frac{r^2}{L^2}\left[- f(r) dt^2 + dx^2 + dy^2\right]+\frac{L^2}{r^2 f(r)} dr^2\, ,\\
A&=\mu\Big(1-\frac{r_0}{r}\Big)dt\, ,\label{gfield} 
\end{align}

\noindent which is a solution of the equations of motion obtained from \eqref{EMaction} with
\bea\label{fmu}
f(r)=1-M\left(\frac{r_0}{r}\right)^3+Q^2\left(\frac{r_0}{r}\right)^4\, , \qquad \qquad \mu=\frac{Qr_0}{L^2}\, .
\eea

\noindent Here $r_0$ is the position of the horizon, given by the largest real root of $f(r_0)=0$, so that $M=1+Q^2$ in our conventions. The black hole temperature takes the form
\bea\label{temp}
T=\frac{r_0}{4\pi L^2}(3-Q^2)\, , 
\eea

\noindent while its entropy, charge and energy densities are given by 
\bea\label{ece}
s=\frac{2\pi}{\kappa_4^2}\left(\frac{r_0}{L}\right)^2\, ,\qquad \qquad \rho=\frac{2}{\kappa_4^2}\left(\frac{r_0}{L}\right)^2Q\, ,\qquad\qquad \epsilon=\frac{r_0^3}{\kappa_4^2L^4}M\, ,
\eea

\noindent respectively \cite{Chamblin:1999tk}. This background is holographically dual to a $(2+1)$-dimensional strongly-coupled field theory at finite temperature $T$ and finite charge density $\rho$. The entropy and energy densities of the dual field theory are given by $s$ and $\epsilon$ in \eqref{ece}, respectively. Similarly, the asymptotic value of the bulk gauge field $A_t(r \to \infty)=\mu$ is interpreted in the dual theory as the chemical potential for the (electric) charge density.  Little is known about the details of the dual theory from the field theory perspective. On the other hand, using holography a lot has been learned (especially thermodynamical properties) regarding its strong-coupling behavior; see \cite{Chamblin:1999tk} and its citations.

 The background becomes extremal when $Q^2=3$; for this value of the charge the temperature is zero, but the entropy density remains finite. Since the solution is invariant under changing the sign of $A_t$ we can choose $\mu$ to be positive, so that in our conventions $Q = \sqrt{3}$ at extremality.  In this paper we will mainly work in the extremal limit, and refer to the corresponding dual theory as the ``boundary field theory". As we will discuss in section \ref{section: Green's functions and criticality}, interesting properties of the boundary field theory in the IR limit stem from the features of the extremal near horizon geometry, which we briefly review below. 
%%%%%%%%%%%%%%%%%%%%%%%%%%%%%%%%%%%%%%%%%%%%%%%%%%%%%%%%%%%%%%%%%%%%%%%%%%%%%%%%%%%%%%%%%%
\subsection{Near horizon geometry at extremality}
Although the black hole temperature vanishes at extremality, its horizon area remains finite, a fact whose dual interpretation is that the boundary theory has a finite entropy density at zero temperature. In the extremal limit, $f(r)$ in the background metric (\ref{metric}) takes the form 
\bea\label{extf}
f(r)=1-4\left(\frac{r_0}{r}\right)^3+3\left(\frac{r_0}{r}\right)^4\, ,
\eea

\noindent which has a double zero at the horizon, and can be approximated near that region (to the leading order in $r-r_0$) by 
\bea
f(r)\simeq \frac{6}{r_0^2}(r-r_0)^2\, .
\eea

\noindent The near horizon geometry at extremality is $\ads_2\times \mathds{R}^2$. To see the emergence of this geometry, first change the radial coordinate $r$ to $\eta$ defined by 
\beq\label{definition eta}
r-r_0=\frac{L^2}{6\eta}\, .
\eeq

\noindent There is then a scaling limit \cite{Faulkner:2009wj} in which
\begin{align}\label{mgnhrads2}
ds^2=\frac{L^2}{6\eta^2}\Big(-dt^2+d\eta^2\Big)+\frac{r_0^2}{L^2}\Big(dx^2+dy^2\Big),\qquad\qquad A&=\frac{Q}{6\eta}dt\, . 
\end{align}

\noindent The curvature radius of the $\ads_2$ factor is $L_2=L/\sqrt{6}$. The radial coordinate is interpreted holographically as the renormalization scale of the dual field theory, and the near horizon region corresponds to its IR limit. This implies that  the $\ads_2\times\mathds{R}^2$ geometry encodes the IR physics ($\omega\to 0$) of the boundary theory.
%%%%%%%%%%%%%%%%%%%%%%%%%%%%%%%%%%%%%%%%%%%%%%%%%%%%%%%%%%%%%%%%%%%%%%%%%%%%%%%%%%%%%%%%%%
\subsection{Linearized equations of motion: master fields}
The holographic calculation of the vector current and energy-momentum tensor operators of the boundary field theory entails solving the linearized Einstein-Maxwell equations for the corresponding fluctuations. We first define
\bea\label{fluct}
g_{\mu\nu}= \bar g_{\mu\nu}+h_{\mu\nu}\, , \qquad A_\mu= \bar A_\mu+ a_\mu\, ,
\eea

\noindent where $\bar g_{\mu\nu}$ and  $\bar A_\mu$ denote the background metric and gauge field, respectively, and $h_{\mu\nu}$ and $a_\mu$ represent the fluctuations. We work in the so-called radial gauge, where 
\bea\label{rgauge}
a_r=0, \qquad h_{r\nu}=0\, ,
\eea

\noindent with $\nu=\{t,x,y,r\}$. We proceed by Fourier transforming the fluctuations
\begin{align}\label{foux}
h_{\mu\nu}(t,x,r)\sim e^{-i\omega t} e^{ikx} h_{\mu\nu}(r)\, ,\qquad
a_\mu(t,x,r)\sim e^{-i\omega t} e^{ik x} a_{\mu}(r)\, ,
\end{align}

\noindent where, without loss of generality, we have used the rotation invariance in the $(x,y)$ plane to set $k_y=0$ and defined $k_x\equiv k$.  The fluctuations split into decoupled sectors depending on whether they are even or odd with respect to parity, $y\to -y$. Accordingly, $h_{ty}$, $h_{xy}$, $a_{y}$ have odd parity while $h_{tt}$, $h_{tx}$, $h_{xx}$, $h_{yy}$, $a_{t}$, $a_{x}$ all have even parity.  The analysis of the odd parity (shear) modes was performed in \cite{Edalati:2010hk}. In this paper, we study the even parity modes which translate into the longitudinal modes of the boundary theory. 

It is convenient to raise indices in the metric fluctuations by the background metric $\bar g_{\mu\nu}$, and work instead with ${h^t}_t$, ${h^x}_t$, ${h^x}_x$, ${h^y}_y$. We also define the dimensionless quantities
\bea\label{dimensionless}
u=\frac{r}{r_0}\, ,\qquad\qquad \wn=\frac{\omega}{\mu}\, , \qquad\qquad \qn=\frac{k}{\mu}\, . 
\eea

\noindent The linearized Maxwell equations read\footnote{We keep the black hole charge $Q$ arbitrary ($0<Q\leq \sqrt{3}$) in writing the equations in this section. In later sections, when we discuss the extremal case, we set $Q=\sqrt{3}$.}
\begin{align}
0 &= 2u^2f(u)\left[u^2a'_t(u)\right]'-2Q^2\qn\left[\qn a_t(u)+\wn a_x(u)\right] \nonumber\\
&\phantom{=}- \mu\, u^2f(u)\left[{h^t}_t'(u)-{h^x}_x'(u)-{h^y}_y'(u)\right],\label{eqt}\\
0 &= u^2f(u)\left[u^2f(u)a'_x(u)\right]'+Q^2\wn\left[\qn a_t(u)+\wn a_x(u)\right]+ \mu\, u^2f(u){h^x}_t'(u)\, ,\label{eqx}\\
0 &= 2u^2\left[\wn a'_t(u)+\qn f(u)a'_x(u)\right]-\mu\, \wn\left[{h^t}_t(u)-{h^x}_x(u)-{h^y}_y(u)\right] +2\mu\, \qn{h^x}_t(u)\, , \label{equ}
\end{align}

\noindent where $f(u)$ is given in (\ref{extf}). Notice that not all of the above equations are independent; equation \eqref{eqx} can be obtained from equations \eqref{eqt} and \eqref{equ}, for example. For the linearized Einstein equations we obtain 
\begin{align}
0 &= f(u)\biggl\{2u^6f(u){h^t}_t''(u)+\left[10u^5 f(u)+3u^6 f'(u)\right]{h^t}_t'(u)\nonumber\\
&\hphantom{= f(u)\biggl\{}  +\left[u^6f'(u)+2u^5f(u)\right]\left[{h^x}_x'(u)+{h^y}_y'(u)\right]\biggr\}-8\mu ^{-1}Q^2u^2f(u)a'_t(u)\\
&\phantom{=}+2Q^2f(u)\left(2-\qn^2u^2\right){h^t}_t(u)+2Q^2\wn^2u^2\left[{h^x}_x(u)+{h^y}_y(u)\right]+4Q^2\qn\wn u^2{h^x}_t(u)\, ,\nonumber\\ \nonumber\\
0 &= f(u)\biggl\{u^6{h^x}_x''(u)+\left[5u^5 f(u)+u^6 f'(u)\right]{h^x}_x'(u)+u^5f(u)\left[{h^t}_t'(u)+{h^y}_y'(u)\right]\biggr\}\nonumber\\
&\phantom{=} + Q^2\wn^2 u^2 {h^x}_x(u)-Q^2f(u)\left(2+\qn^2u^2\right){h^t}_t(u)-Q^2\qn^2u^2f(u){h^y}_y(u)\\
&\phantom{=}+ 2Q^2\qn\wn u^2{h^x}_t(u) +4\mu^{-1}Q^2u^2f(u)a'_t(u)\, ,\nonumber\\ \nonumber \\
0 &= f(u)\biggl\{u^6{h^y}_y''(u)+\left[5u^5 f(u)+u^6 f'(u)\right]{h^y}_y'(u)+u^5f(u)\left[{h^t}_t'(u)+{h^x}_x'(u)\right]\biggr\}\nonumber\\
&\phantom{=}-2Q^2f(u){h^t}_t(u)+Q^2u^2\left(\wn^2-f(u)\qn^2\right){h^y}_y(u)+4\mu^{-1}Q^2u^2f(u)a'_t(u)\, ,\\ \nonumber\\
0 &= f(u)\left[u^6 {h^x}_t''(u)+4u^5 {h^x}_t'(u)\right]+Q^2\qn\wn u^2 {h^y}_y(u)+4\mu^{-1}Q^2u^2f(u)a'_x(u)\, , \\ \nonumber\\
0 &= 2u^6f(u)\left[{h^t}_t''(u)+{h^x}_x''(u)+{h^y}_y''(u)\right]+\left[u^6f(u)\right]'\left[{h^t}_t'(u)+{h^x}_x'(u)+{h^y}_y'(u)\right]\nonumber\\
&+2u^6f'(u){h^t}_t'(u)+4Q^2{h^t}_t(u)-8\mu^{-1}Q^2u^2a'_t(u)\, ,\\ \nonumber\\
0 &= 2\wn f(u)\left[{h^x}_x'(u)+{h^y}_y'(u)\right]+2\qn f(u) {h^x}_t'(u) -\wn f'(u)\left[{h^x}_x(u)+{h^y}_y(u)\right]\\
&\phantom{=}-2\qn f'(u){h^x}_t(u)\, ,\nonumber\\  \nonumber\\
0 &= 2\qn u^4f(u)\left[{h^t}_t'(u)+{h^y}_y'(u)\right]-2\wn u^4 {h^x}_t'(u)+\qn u^4 f'(u){h^t}_t(u)\nonumber\\
&\phantom{=}-8\mu^{-1}Q^2\left[\qn a_t(u)+\wn a_x(u)\right]\, . \label{equx}  
\end{align}

The linearized Einstein-Maxwell equations presented above can be reduced to a set of two decoupled second-order differential equations for the so-called ``master fields" $\Phi_\pm(u)$, introduced by Ishibashi and Kodama in \cite{Kodama:2003kk}.  In appendix \ref{appendix: master fields} we explain in detail how to obtain the decoupled equations for the master fields. Basically, one defines the gauge-invariant quantities
\begin{align}\label{calA}
{\cal A}(u)&=u^2a'_t(u)+\mu {h^y}_y(u)-\frac{1}{2}\mu{h^t}_t(u),\\
\Phi(u)&=u{h^y}_y(u)-\left[Q^2\qn^2+u^3f'(u)\right]^{-1}u^4 f(u)\left[{h^x}_x'(u)+{h^y}_y'(u)\right],\label{Phi}
\end{align}

\noindent in terms of which the equations \eqref{eqt}--\eqref{equx} reduce to a set of two coupled second-order differential equations given by \eqref{eqA} and \eqref{eqPhi}. The resulting two equations can be decoupled by introducing the master fields\footnote{The master fields defined in \eqref{Phipmext} differ from the ones given in \cite{Kodama:2003kk} by a multiplicative constant.}
\begin{align}\label{Phipmext}
\Phi_\pm(u)=\alpha_{\pm}(u) \Phi(u) +\mu^{-1}Q^2{\cal A}(u)\, , 
\end{align}

\noindent where we have defined
\begin{align}
\alpha_{\pm}(u)= \frac{1}{2}F_\pm(Q)-\frac{Q^2}{u}\, , \qquad F_\pm(Q)=\frac{3}{4}\Big[(1+Q^2)\pm\sqrt{(1+Q^2)^2+(16/9)Q^4\qn^2}\Big].
\end{align}

\noindent The decoupled equations for the master fields then take the form
\begin{align}\label{sphipmeq}
u^2f(u)\left[u^2f(u)\Phi'_\pm(u)\right]'+\left[ Q^2\wn^2-U_{\pm}(u)\right]\Phi_\pm(u)=0\, ,
\end{align}

\noindent with 
\begin{align}\label{extremalUpm}
U_{\pm}(u)&=\pm \gamma f(u)\left\{\alpha_{\pm}(u)V(u)+\frac{Q^2}{4\alpha_{\pm}(u)}\left[Q^2\qn^2H(u)+8Q^2f(u)\right]+\frac{2Q^2}{u}\left[H(u)-2Q^2\qn^2\right]\right. \nonumber\\ 
&~~~~~~~~~~~~~~~~\left. +\frac{Q^2}{H(u)}\left[48u^3f(u)(f(u)-1)+20u^4f(u)f'(u)\right]\right\},
\end{align}

\noindent where $\gamma$, $V(u)$ and $H(u)$ in \eqref{extremalUpm} have been defined in appendix \ref{appendix: master fields}. In the special case where the (electric) charge of the black hole background vanishes, namely $Q=0$, the electromagnetic and gravitational perturbations decouple, and the master fields introduced above reduce to a single field carrying the information from the Einstein sector. In our notation, this field is given $\Phi_{+}(u)=\Phi(u)$, and it satisfies equation \eqref{ucmastereq}. The gauge sector is dealt with by introducing a second independent gauge-invariant mode. 

In the next section, we will see that the Ishibashi-Kodama master fields introduced in \eqref{Phipmext} are not well suited to the analysis of the sound-channel retarded correlators of the boundary theory via holography.\footnote{They are a convenient choice, however, to analyze the shear-channel correlators \cite{Edalati:2010hk}.} As we explain below, this is because of the relationship between the leading term in the asymptotic expansion of $\Phi_\pm(u)$ and those of the metric and gauge field fluctuations. The guiding principle is that, in gauge/gravity duality, the ambiguity in the choice of asymptotic boundary conditions for the gauge-invariant fields is removed by demanding the quasinormal mode spectrum in the gravity side to coincide with the poles of the retarded correlators in the dual field theory \cite{Kovtun:2005ev}. 

 The observation that the Ishibashi-Kodama-like master fields with asymptotic Dirichlet boundary conditions may not be appropriate for the analysis of the sound-channel correlators via holography was first made in \cite{Michalogiorgakis:2006jc}, for the case of an uncharged AdS$_4$ black hole. In order to obtain the sound mode in the boundary theory, the authors of \cite{Michalogiorgakis:2006jc} proposed that one should instead impose a Robin (``mixed") boundary condition on the master field (recall that in the case where the background is uncharged, $\Phi_{\pm}$ reduce to a single field which carries the information about the gravity fluctuations). Our approach here is different: we introduce new master fields by taking linear combinations of $\Phi_\pm(u)$ and their derivatives, such that the leading term in the asymptotic expansion of the new master fields contains the leading term in the asymptotic expansions of the metric and gauge field fluctuations only  (which are the sources for  the corresponding operators in the boundary theory). In other words, we perform a canonical transformation in the space of master fields and their conjugate momenta, so that the new master fields satisfy a Dirichlet boundary condition asymptotically. The advantage of this approach is twofold: on one hand it makes the relation between the asymptotics of the master fields and the gauge-variant sources more transparent, while at the same time it allows us to use the various known techniques for the calculation of the quasinormal mode spectrum, such as the (Leaver's) matrix method we will employ later on.
%%%%%%%%%%%%%%%%%%%%%%%%%%%%%%%%%%%%%%%%%%%%%%%%%%%%%%%%%%%%%%%%%%%%%%%%%%%%%%%%%%%%%%%%%%
%%%%%%%%%%%%%%%%%%%%%%%%%%%%%%%%%%%%%%%%%%%%%%%%%%%%%%%%%%%%%%%%%%%%%%%%%%%%%%%%%%%%%%%%%%
\section{Green's functions and criticality}\label{section: Green's functions and criticality}
In this section we present the asymptotic expansions of various bulk fields and the renormalized boundary action, and use them to obtain formal expressions for the retarded two-point functions of the vector current and energy-momentum tensor operators of the boundary theory in the sound channel. We then analyze these expressions in the small frequency limit and comment on the emergent critical behavior stemming from the properties of the near horizon geometry of the extremal background.
%%%%%%%%%%%%%%%%%%%%%%%%%%%%%%%%%%%%%%%%%%%%%%%%%%%%%%%%%%%%%%%%%%%%%%%%%%%%%%%%%%%%%%%%%%
\subsection{Asymptotic expansion of the bulk modes}
Asymptotically as $u\to \infty$, the solution of the Einstein-Maxwell equations \eqref{eqt}--\eqref{equx} have the expansions
\begin{align}\label{amuexpan}
a_\mu(u\to\infty)&={\hat a}_\mu\left[1+{\cal O}\left(u^{-2}\right)\right]+\frac{1}{u}\pi_\mu\left[1+{\cal O}\left(u^{-2}\right)\right],\\
{h^\mu}_\nu(u\to\infty)&= {\hat h^\mu}_{~\nu}\left[1+{\cal O}\left(u^{-2}\right)\right]+\frac{1}{u^3}{\pi^\mu}_{\nu}\left[1+{\cal O}\left(u^{-2}\right)\right] ,\label{hmunuexpan}
\end{align}

\noindent where it is understood that ${\hat a}_\mu$, $\pi_\mu$,  ${\hat h^\mu}_{~\nu}$ and ${\pi^\mu}_{\nu}$, hence $a_\mu(u)$ and ${h^\mu}_\nu(u)$, are all functions of $\wn$ and $\qn$. We have denoted the leading constant terms in the asymptotic expansions of the modes by a hat. These are the sources for the corresponding operators in the boundary theory. Inserting \eqref{amuexpan} and \eqref{hmunuexpan} into the six second-order Einstein-Maxwell equations, we find 
\begin{align}
a_x(u\to\infty)&={\hat a}_x+\frac{1}{u}\pi_x-\frac{Q^2}{2u^2}\wn\left(\qn {\hat a}_t+\wn {\hat a}_x \right)-\frac{Q^2}{6u^3}\wn(\qn \pi_t +\wn \pi_x-\qn {\hat h^y}_{~y})\nonumber\\
&~~~+\frac{1}{24u^4}\Big[6 {\pi^x}_t+6\left(1+Q^2\right)\pi_x+Q^4\wn\left(\wn^2-\qn^2\right)\left(\qn {\hat a}_t+\wn {\hat a}_x\right)\Big]+\cdots\, ,
\end{align}
\begin{align}\label{atasym}
a_t(u\to\infty)&={\hat a}_t+\frac{1}{u}\pi_t+\frac{Q^2}{2u^2}\qn\left(\qn {\hat a}_t+\wn {\hat a}_x \right)+\frac{Q^2}{24u^3}\Big[4\qn(\qn \pi_t +\wn \pi_x-\qn {\hat h^y}_{~y})\nonumber\\
&~~~-\left(2\wn\qn{\hat h^x}_{~t}+\wn^2{\hat h^x}_{~x}-\qn^2{\hat h^t}_{~t}+\left(\wn^2-\qn^2\right){\hat h^y}_{~y}\right)\Big]\nonumber\\ 
&~~~+\frac{1}{24u^4}\Big[3\left({\pi^x}_x+{\pi^y}_y-{\pi^t}_t\right)+Q^4\qn\left(\qn^2-\wn^2\right)\left(\qn {\hat a}_t+\wn {\hat a}_x\right)\Big]+\cdots\, , \\\nonumber\\
{h^t}_t(u\to\infty)&={\hat h^t}_{~t}+\frac{Q^2}{4u^2}\left[2\wn\qn{\hat h^x}_{~t}+\wn^2{\hat h^x}_{~x}-\qn^2{\hat h^t}_{~t}+\left(\wn^2+\qn^2\right){\hat h^y}_{~y}\right]+\frac{1}{u^3}{\pi^t}_t\nonumber\\
-\frac{Q^2}{16u^4}\Big[&24\pi_t+12{\hat h^t}_{~t}-Q^2\qn^2\left(2\wn\qn{\hat h^x}_{~t}+\wn^2{\hat h^x}_{~x}-\qn^2{\hat h^t}_{~t}-\left(\wn^2-\qn^2\right){\hat h^y}_{~y}\right)\Big]+\cdots\, ,\label{httasym}\\ \nonumber\\
{h^x}_x(u\to\infty)&={\hat h^x}_{~x}+\frac{Q^2}{4u^2}\left[2\wn\qn{\hat h^x}_{~t}+\wn^2{\hat h^x}_{~x}-\qn^2{\hat h^t}_{~t}-\left(\wn^2+\qn^2\right){\hat h^y}_{~y}\right]+\frac{1}{u^3}{\pi^x}_x\nonumber\\
+\frac{Q^2}{16u^4}&\Big[8\pi_t+4{\hat h^t}_{~t}-Q^2\wn^2\left(2\wn\qn{\hat h^x}_{~t}+\wn^2{\hat h^x}_{~x}-\qn^2{\hat h^t}_{~t}-\left(\wn^2-\qn^2\right){\hat h^y}_{~y}\right)\Big]+\cdots\, ,\label{hxxasym}\\\nonumber\\
{h^y}_y(u\to\infty)&={\hat h^y}_{~y}-\frac{Q^2}{4u^2}\left[2\wn\qn{\hat h^x}_{~t}+\wn^2{\hat h^x}_{~x}-\qn^2{\hat h^t}_{~t}-\left(\wn^2-\qn^2\right){\hat h^y}_{~y}\right]+\frac{1}{u^3}{\pi^y}_y\nonumber\\
+\frac{Q^2}{16u^4}\Big[8\pi_t + &4{\hat h^t}_{~t}+Q^2\left(\wn^2-\qn^2\right)\left(2\wn\qn{\hat h^x}_{~t}+\wn^2{\hat h^x}_{~x}-\qn^2{\hat h^t}_{~t}-\left(\wn^2-\qn^2\right){\hat h^y}_{~y}\right)\Big]+\cdots\, ,\label{hyyasym}\\ \nonumber \\
{h^x}_t(u\to\infty)&={\hat h^x}_{~t}+\frac{Q^2}{4u^2}\left(2\wn\qn{\hat h^y}_{~y}\right)+\frac{1}{u^3}{\pi^x}_t\nonumber\\
+&\frac{Q^2}{16u^4}\Big[16\pi_x+Q^2\wn\qn\left(2\wn\qn{\hat h^x}_{~t}+\wn^2{\hat h^x}_{~x}-\qn^2{\hat h^t}_{~t}-\left(\wn^2-\qn^2\right){\hat h^y}_{~y}\right)\Big]+\cdots\, .
\end{align}

\noindent When expanded asymptotically, the four constraint equations (the $uu$, $ut$, and $ux$-components of the Einstein equations together with the $u$-component of the Maxwell equations) yield
\begin{align}\label{conspitpix}
\qn\pi_x+\wn\pi_t&=\qn{\hat h^x}_{~t}+\frac{1}{2}\wn\left({\hat h^x}_{~x}+{\hat h^y}_{~y}-{\hat h^t}_{~t}\right),\\
\qn {\pi^y}_y+\qn {\pi^t}_t-\wn{\pi^x}_t&=\frac{1}{2}\left(1+Q^2\right)\qn {\hat h^t}_{~t}-\frac{4Q^2}{3}\left(\qn {\hat a}_t+\wn {\hat a}_x \right),\\
\qn {\pi^x}_t+\wn{\pi^x}_x+\wn{\pi^y}_y&=-\frac{1}{2}\left(1+Q^2\right)\left(2\qn{\hat h^x}_{~t}+\wn{\hat h^x}_{~x}+\wn{\hat h^y}_{~y}\right),\\
{\pi^t}_t+{\pi^x}_x+{\pi^y}_y&=0\, .\label{conspittpiyy}
\end{align}
\noindent Using \eqref{sphipmeq}, the asymptotic expansion of the master fields $\Phi_{\pm}$ takes the form 
\begin{align}
\Phi_{\pm}(u&\to \infty)=\Phi^{(0)}_\pm\left\{1+\frac{1}{2u^2}\left[\frac{4F^2_\pm}{Q^4\qn^4}+Q^2\left(\qn^2-\wn^2\right)\right]-\frac{F_\pm}{3Q^2\qn^2u^3}\left[\frac{16F^2_\pm}{Q^4\qn^4}+Q^2\qn^2\right]+\cdots\right\}\nonumber\\
&+\frac{\Phi^{(1)}_\pm}{u}\left\{1+\frac{1}{6u^2}\left[\frac{8F^2_\pm}{Q^4\qn^4}+Q^2\left(\qn^2-\wn^2\right)\right]-\frac{1}{6F_\pm u^3}\left[\frac{16F^4_\pm}{Q^6\qn^6}+Q^4\qn^2\right]+\cdots\right\}.
\end{align}

\noindent From the definition of the master fields in \eqref{Phipmext} and the asymptotic expansions of  $a_t(u)$,  ${h^t}_t(u)$, ${h^x}_x(u)$ and ${h^y}_y(u)$ given in \eqref{atasym}, \eqref{httasym}, \eqref{hxxasym} and \eqref{hyyasym}, respectively, one concludes that 
\begin{align}
\Phi^{(0)}_{\pm} &= \frac{1}{Q^2\qn^2}\left(-\frac{3}{2}F_{\pm}{\pi^t}_t-\mu^{-1}Q^4\qn^2\pi_t+F^2_{\pm}{\hat h^y}_{~y}-\frac{1}{2}Q^4\qn^2{\hat h^t}_{~t}\right),\label{hatPhipm}\\
\Phi^{(1)}_{\pm} &= -\frac{2F_\pm}{Q^4\qn^4}\left(-\frac{3}{2}F_{\pm}{\pi^t}_t-\mu^{-1}Q^4\qn^2\pi_t+F^2_{\pm}{\hat h^y}_{~y}-\frac{1}{2}Q^4\qn^2{\hat h^t}_{~t}\right)\nonumber\\
&\phantom{=}\,\, -\frac{1}{4}F_\pm \hat z_h-\mu^{-1}Q^2\hat z_a\, ,\label{PhipmPipm}
\end{align}

\noindent where we have defined
\begin{align}
{\hat z}_a&=Q^2\qn\left(\qn {\hat a}_t+\wn {\hat a}_x\right),\label{za}\\
{\hat z}_h&=Q^2\left[2\wn\qn{\hat h^x}_{~t}+\wn^2{\hat h^x}_{~x}-\qn^2{\hat h^t}_{~t}+\left(\qn^2-\wn^2\right){\hat h^y}_{~y}\right].\label{zh}
\end{align}

\noindent Notice the appearance of the momenta ${\pi^t}_t$ and $\pi_t$ in $\Phi^{(0)}_{\pm}$: this is the reason why the Ishibashi-Kodama master fields $\Phi_{\pm}(u)$ are not appropriate for the holographic analysis of the boundary theory Green's functions in the sound channel. As we mentioned before, this issue is specific to the sound channel and does not arise in the shear channel. 
%%%%%%%%%%%%%%%%%%%%%%%%%%%%%%%%%%%%%%%%%%%%%%%%%%%%%%%%%%%%%%%%%%%%%%%%%%%%%%%%%%%%%%%%%%
\subsection{New master fields}
From equations \eqref{hatPhipm} and \eqref{PhipmPipm} we obtain
\begin{align}
\frac{2F_\pm}{Q^2\qn^2}\Phi^{(0)}_{\pm}+\Phi^{(1)}_{\pm}=-\frac{1}{4}F_\pm{\hat z}_h-\mu^{-1}Q^2{\hat z}_a\, ,
\end{align}

\noindent which is independent of the momenta  ${\pi^t}_t$ and $\pi_t$. This suggests that we define two new master fields 
\begin{align}\label{Psipm}
\Psi_\pm(u)=\frac{2F_\pm}{Q^2\qn^2}\Phi_\pm(u)-u^2f(u)\Phi'_\pm(u)\, ,
\end{align}

\noindent whose leading asymptotic (constant) terms do not depend on the field momenta. Using the fact that $\Phi_{\pm}(u)$ satisfy \eqref{sphipmeq}, the decoupled differential equations for $\Psi_\pm(u)$ take the form
\begin{align}\label{equation for Psi}
\left[u^2f(u)\Psi^{\prime}_\pm(u)\right]'+J_\pm(u)\Psi^{\prime}_\pm(u)+K_\pm(u)\Psi_\pm(u)=0\, ,
\end{align}
where
\begin{align}\label{definition J and K}
J_\pm(u)&=\left[\frac{4F^2_\pm}{Q^4\qn^4}+Q^2\wn^2-U_\pm(u)\right]^{-1}u^2f(u)U'_\pm(u)\, ,\\
K_\pm(u)&=\frac{1}{u^2f(u)}\left[Q^2\wn^2+\frac{2F_\pm}{Q^2\qn^2}J_\pm(u)-U_\pm(u)\right].\label{definition J and K 2}
\end{align}

\noindent Asymptotically, $\Psi_\pm(u)$ take the form
\begin{align}\label{asymptotics of Psipm}
\Psi_\pm(u\to\infty)&=\Psi^{(0)}_\pm\left[1+{\cal O}\left(u^{-2}\right)\right]+\frac{1}{u}\Psi^{(1)}_\pm\left[1+{\cal O}\left(u^{-1}\right)\right].
\end{align}

\noindent Using \eqref{Psipm}, \eqref{hatPhipm} and \eqref{PhipmPipm},  the two coefficients $\Psi^{(0)}_\pm$ and $\Psi^{(1)}_\pm$ are related to the asymptotic values (sources and momenta) of the metric and gauge field fluctuations through
\begin{align}
\hat \Psi_{\pm}&\equiv\Psi^{(0)}_\pm=-\frac{1}{4}F_\pm{\hat z}_h-\mu^{-1}Q^2{\hat z}_a\, ,\label{Psihat}\\
\Psi^{(1)}_{\pm} &= \frac{C_\pm}{Q^2\qn^2}\left(-\frac{3}{2}F_{\pm}{\pi^t}_t-\mu^{-1}Q^4\qn^2\pi_t+F^2_{\pm}{\hat h^y}_{~y}-\frac{1}{2}Q^4\qn^2{\hat h^t}_{~t}\right)\nonumber\\
&\phantom{=}-\frac{2F_\pm}{Q^2\qn^2}\left(\frac{1}{4}F_\pm \hat z_h+\mu^{-1}Q^2\hat z_a\right),\label{PsihatPihat}
\end{align}

\noindent where $\hat z_a$ and $\hat z_h$ are given by \eqref{za} and \eqref{zh}, and we have defined 
\begin{align}
C_\pm=\frac{4F^2_\pm}{Q^4\qn^4}+Q^2\left(\qn^2-\wn^2\right).
\end{align}

\noindent In order to compute retarded correlators, the Lorentzian AdS/CFT recipe of \cite{Son:2002sd} instructs us to apply infalling boundary conditions for the bulk fields at the black hole horizon. Imposing infalling boundary conditions for $\Psi_\pm(u)$ at the horizon, $\Psi^{(1)}_{\pm}$ will be related to $\hat \Psi_{\pm}$. Parameterizing this relationship as
\begin{equation}\label{momenta in terms of source}
\Psi^{(1)}_{\pm}=\left[C_{\pm}\Pi_{\pm} +\frac{2 F_{\pm}}{Q^{2}\qn^{2}}\right]\hat \Psi_{\pm}\, ,
\end{equation}

\noindent equations \eqref{Psihat} and \eqref{PsihatPihat} reduce to 
\begin{align}
Q^2\qn^2\Pi_\pm\hat \Psi_{\pm}&=-\frac{3 }{2} F_\pm {\pi^t}_t- \mu^{-1}Q^4\qn^2 \pi_t-\frac{1}{2}Q^4\qn^2{\hat h^t}_{~t}+ F^2_\pm{\hat h^y}_{~y}\, .\label{PsihatPihatrelation}
\end{align}

\noindent Solving the above two equation for ${\pi^t}_t$ and $\pi_t$, we obtain
\begin{align}\label{pitt}
{\pi^t}_t&=\left(1+Q^2\right){\hat h^y}_{~y}+\frac{1}{6}\frac{Q^2\qn^2}{F_{+}-F_{-}}\Big[\left(F_{+}\Pi_{+}-F_{-}\Pi_{-}\right)\hat z_h+4\mu^{-1}Q^2\left(\Pi_{+}-\Pi_{-}\right)\hat z_a\Big],\\
\pi_t&=\mu{\hat h^y}_{~y}-\frac{1}{2}\mu{\hat h^t}_{~t}+\frac{1}{4}\frac{1}{F_{+}-F_{-}}\Big[\mu Q^2\qn^2\left(\Pi_{+}-\Pi_{-}\right)\hat z_h-4\left(F_{-} \Pi_{+}-F_{+} \Pi_{-}\right)\hat z_a\Big].\label{pit}
\end{align}
$\Pi_\pm$ will be determined numerically. The other four field momenta are easily obtained by substituting equations \eqref{pitt} and \eqref{pit} into equations \eqref{conspitpix}--\eqref{conspittpiyy}. 
%%%%%%%%%%%%%%%%%%%%%%%%%%%%%%%%%%%%%%%%%%%%%%%%%%%%%%%%%%%%%%%%%%%%%%%%%%%%%%%%%%%%%%%%%%
\subsection{Boundary action and retarded Green's functions}
The renormalized action in our case is given by \cite{Kraus:1999di}
\begin{align}\label{action}
2\kappa_{4}^{2}\,S_{\rm ren} &=\int d^{4}x \sqrt{-g}\left(R+\frac{6}{L^2}- L^{2}F_{\mu\nu}F^{\mu\nu}\right)\nonumber\\
&-\int_{\partial M}d^{3}x\,\sqrt{|\gamma|}2K
- \frac{4}{L}\int_{\partial M}d^{3}x\,\sqrt{|\gamma|} - L\int_{\partial M}d^{3}x\,\sqrt{|\gamma|}{~^{(3)}R}\, ,
\end{align}

\noindent where $\gamma_{\mu\nu}$,  $K$ and ${^{(3)}R}$ are, respectively, the induced metric, the trace of the second fundamental form and the intrinsic curvature of the 3-dimensional boundary $\partial M$.\footnote{If $n^{\mu}$ denotes the components of the outward pointing unit normal vector to the (timelike) boundary, we have $\gamma_{\mu\nu} = g_{\mu\nu} -n_{\mu}n_{\nu}$ and $K_{\mu\nu} = -(1/2)\pounds_{n}\gamma_{\mu\nu} =-\gamma^{\rho}_{\phantom{\rho}\mu}\nabla_{\rho}n_{\nu} \quad \Rightarrow \quad K = \gamma^{\mu\nu}K_{\mu\nu}=-\nabla_{\mu}n^{\mu}$ .} The counterterm proportional to ${^{(3)}R}$ is introduced in order to remove a linear divergence in the on-shell action for the RN-AdS$_{4}$ background. To obtain Green's functions holographically, it suffices to  consider the (gravity) action up to quadratic terms in the fluctuations. The renormalized on-shell action then takes the form 
\begin{equation}
S_{\rm ren} = S^{(1)}_{\rm ren}+ S^{(2)}_{\rm ren}\, ,
\end{equation}

\noindent where the $S^{(1)}_{\rm ren}$ piece is linear in fluctuations and determines the one-point functions of the dual operators, while $S^{(2)}_{\rm ren}$ is quadratic in fluctuations and determines the two-point functions of the conserved currents in the dual field theory. In our case, $S^{(1)}_{\rm ren}$ is worked out to be
\begin{equation}
S^{(1)}_{\rm ren} = \frac{1}{2}\int_{\partial M}d^{3}x\,\langle T^{j}_{\phantom{j}i}\rangle \hat{h}^{i}_{\phantom{i}j} + \int_{\partial M}d^{3}x\,\langle J^{t}\rangle \hat{a}_{t}\, ,
\end{equation}

\noindent with the one-point functions given by
\begin{align}\label{one point function stress tensor}
\langle T^{i}_{\phantom{i}j}\rangle &= \lim_{u\rightarrow \infty}\frac{1}{\kappa_{4}^{2}}\sqrt{|\bar{\gamma}|}\left[\bar{K}^{i}_{\phantom{i}j} - \left(\bar{K}   +  \frac{2}{L}\right)\delta^{i}_{\phantom{i}j}\right], \\
 \langle J^{t}\rangle &=  \lim_{u\rightarrow \infty}\frac{2L^{2}}{r_{0}\kappa_{4}^{2}}\sqrt{|\bar{g}|}\left(\partial_{u}\bar{A}_{t}\right),\label{one point function current}
\end{align}

\noindent where the bar denotes quantities that are evaluated in the background and $i,j = t,x,y$. In the same way, $S^{(2)}_{\rm ren}$ is given by
\begin{align}\label{onshellactioneu}
S^{(2)}_{\rm ren}=C\lim_{u\rightarrow \infty}\int_{\partial M}d^{3}x &\left\{\frac{u^{4}f}{2}\left[({h^{t}}_{t} + {h^{x}}_{x}){h^{y}}'_{y} + ({h^{t}}_{t} + {h^{y}}_{y}){h^{x}}'_{x} + ({h^{x}}_{x} + {h^{y}}_{y}){h^{t}}'_{t} \right. \right.\nonumber\\
&\left. \qquad \quad \left. + \frac{2}{f}{h^{x}}_{t}{h^{x}}'_{t} \right]+ 4\frac{Q^{2}}{\mu^{2}}u^{2}\left(a_{t}a'_{t} - fa_{x}a'_{x}\right) + {\cal L}_{\rm cont}\right\},
\end{align}

\noindent where  $C={r_0^3}/({4\kappa_4^2L^4})$ and the ``contact term Lagrangian"  ${\cal L}_{\rm cont}$ is given by
\begin{align}
{\cal L}_{\rm cont} &= Q^{2}\frac{u}{\sqrt{f}}{h^{y}}_{y}\left(2\frac{\partial_{x}\partial_{t}}{\mu^{2}}{h^{x}}_{t} - \frac{\partial_{t}^{2}}{\mu^{2}}{h^{x}}_{x} + f\frac{\partial_{x}^{2}}{\mu^{2}}{h^{t}}_{t}\right)-4\frac{Q^{2}}{\mu}\left[a_{x}{h^{x}}_{t} + a_{t}\left({h^{t}}_{t} - \frac{h}{2}\right)\right]\nonumber \\
&\phantom{=}-u^{3}f \left[\left(1-\frac{1}{\sqrt{f}}\right){h^{t}}_{t}{h^{t}}_{t} + 2\left(-1 + \frac{1}{\sqrt{f}} - \frac{uf'}{8f}\right){h^{t}}_{t}\left({h^{x}}_{x}  + {h^{y}}_{y}\right)\right. \nonumber\\
&\left. \phantom{=}\,+\left(1-\frac{1}{\sqrt{f}} + \frac{uf'}{4f}\right)\left( {h^{x}}_{x}{h^{x}}_{x}  + {h^{y}}_{y} {h^{y}}_{y}\right) +2\left(-1 + \frac{1}{\sqrt{f}} - \frac{uf'}{4f}\right){h^{x}}_{x}{h^{y}}_{y} \right. \nonumber\\
 &\left.\phantom{=}+\frac{4}{f}\left(  - 1+\frac{1}{\sqrt{f}}\right){h^{x}}_{t}{h^{x}}_{t}\right].  
\end{align}

\noindent In momentum space we write
\begin{align} \label{onshellfourier}
S^{(2)}_{\rm ren}=-\frac{3}{2}\mu^2C\int dy \int \frac{d\wn d\qn}{(2\pi)^2}&\Big[{\pi^{x}}_x~{\hat h^t}_{~t} + {\pi^{y}}_y~{\hat h^t}_{~t} + {\pi^{t}}_t~{\hat h^x}_{~x}+{\pi^{y}}_y~{\hat h^x}_{~x}+ {\pi^{t}}_t~{\hat h^y}_{~y} \nonumber\\
&+{\pi^{x}}_x~{\hat h^y}_{~y}+2{\pi^{x}}_t~ {\hat h^x}_{~t}+\frac{8Q^2}{3\mu^{2}}\left(\pi_t~\hat a_t-\pi_x~\hat a_x\right)+\hat{{\cal L}}_{\rm cont}\Big],
\end{align}

\noindent where $\hat{{\cal L}}_{\rm cont}$ denotes the Fourier transform of the contact term Lagrangian, and it is understood that ${\pi^{x}}_x~{\hat h^t}_{~t} \equiv {\pi^{x}}_x(\wn,\qn)~{\hat h^t}_{~t} (-\wn,-\qn)$ and so forth. Defining 
\begin{align}
G_1(\wn,\qn)&=C\frac{Q^4}{F_{+}-F_{-}}(\wn^{2} - \qn^{2})^{2}\left(F_{+} \Pi_{+}-F_{-} \Pi_{-}\right)+3C(1 + Q^{2})\left(2 - \frac{\wn^{2}}{\qn^{2}}\right),\label{G1}\\
G_2(\wn,\qn)&=-\frac{8C}{\mu^{2}}\frac{Q^4}{F_{+}-F_{-}}(\wn^{2} - \qn^{2})\left(F_{-} \Pi_{+}-F_{+} \Pi_{-}\right),\label{G2}
\end{align}

\noindent we can write the various retarded Green's functions in a compact way. For notational simplicity, the contact term contributions are not displayed in the expressions below; their explicit form is given in appendix \ref{appendix: contact}. Using the Lorentzian AdS/CFT recipe of \cite{Son:2002sd,Son:2006em}, for the retarded two-point functions of the form $\langle T_{\mu\nu}T_{\alpha\beta}\rangle$ we find 
\begin{align}
G_{tt,tt}(\wn,\qn)&=\frac{1}{2}\frac{\qn^4}{(\wn^{2} - \qn^{2})^{2}}G_1(\wn,\qn)\, , &  G_{xx,tt}(\wn,\qn)&=\frac{1}{2}\frac{\qn^2\wn^{2}}{(\wn^{2} - \qn^{2})^{2}}G_1(\wn,\qn) \, , \label{Gtttt}\\
G_{yy,tt}(\wn,\qn)&=-\frac{1}{2}\frac{\qn^2}{\wn^{2} - \qn^{2}}G_1(\wn,\qn) \, ,& G_{xt,tt}(\wn,\qn)&=-\frac{1}{2}\frac{\qn^{3}\wn}{(\wn^{2} - \qn^{2})^{2}}G_1(\wn,\qn) \, ,\\
G_{xx,xx}(\wn,\qn)&=\frac{1}{2}\frac{\wn^{4}}{(\wn^{2} - \qn^{2})^{2}}G_1(\wn,\qn) \, ,& G_{yy,xx}(\wn,\qn)&=-\frac{1}{2}\frac{\wn^{2}}{\wn^{2} - \qn^{2}}G_1(\wn,\qn) \, ,\\ 
G_{xt,xx}(\wn,\qn)&=-\frac{1}{2}\frac{\qn\wn^{3}}{(\wn^{2} - \qn^{2})^{2}}G_1(\wn,\qn) \, ,& G_{yy,yy}(\wn,\qn)&=\frac{1}{2}G_1(\wn,\qn) \, ,\\
G_{xt,yy}(\wn,\qn)&=\frac{1}{2}\frac{\qn\wn}{\wn^{2} - \qn^{2}}G_1(\wn,\qn)\, ,& G_{xt,xt}(\wn,\qn)&=\frac{\qn^{2}\wn^{2}}{(\wn^{2} - \qn^{2})^{2}}G_1(\wn,\qn) \, .
\end{align}

\noindent Similarly, the current-current correlators of the form $\langle J_{\mu}J_{\nu}\rangle$ read
\begin{align}
G_{t,t}(\wn,\qn)&=\frac{\qn^{2}}{\wn^{2} - \qn^{2}}G_{2}(\wn,\qn)\, ,\\
G_{x,x}(\wn,\qn)&=\frac{\wn^{2}}{\wn^{2} - \qn^{2}}G_{2}(\wn,\qn)\, ,\\
G_{x,t}(\wn,\qn)&=-\frac{\qn\wn}{\wn^{2} - \qn^{2}}G_{2}(\wn,\qn)\, .
\end{align}

\noindent Defining  
\begin{equation}
G_{\mbox{\tiny{mix}}}(\wn,\qn) \equiv \frac{4Q^{2}}{3(1+Q^{2})}\left[\frac{1}{\mu}G_1(\wn,\qn) - \frac{(\wn^{2} - \qn^{2})}{8}\mu G_2(\wn,\qn)\right] ,
\end{equation}

\noindent (recall that we set $Q^{2}=3$ at extremality) the ``mixed" correlators $\langle J_{\mu}T_{\nu\rho} \rangle$ are given by
\begin{align}
G_{t,tt}(\wn,\qn)&=-\frac{\qn^{2}}{(\wn^{2} - \qn^{2})^{2}}G_{\mbox{\tiny{mix}}}(\wn,\qn) \, ,& G_{x,tt}(\wn,\qn)&=\frac{\qn^{3}\wn}{(\wn^{2} - \qn^{2})^{2}}G_{\mbox{\tiny{mix}}}(\wn,\qn) \, ,\\
G_{t,xx}(\wn,\qn)&=-\frac{\qn^{2}\wn^{2}}{(\wn^{2} - \qn^{2})^{2}}G_{\mbox{\tiny{mix}}}(\wn,\qn) \, ,& G_{x,xx}(\wn,\qn)&=\frac{\qn\wn^{3}}{(\wn^{2} - \qn^{2})^{2}}G_{\mbox{\tiny{mix}}}(\wn,\qn)\, ,\\
G_{t,yy}(\wn,\qn)&=\frac{\qn^{2}}{\wn^{2} - \qn^{2}}G_{\mbox{\tiny{mix}}}(\wn,\qn) \, ,& G_{x,yy}(\wn,\qn)&=-\frac{\qn\wn}{\wn^{2} - \qn^{2}} G_{\mbox{\tiny{mix}}}(\wn,\qn)\, ,\\
G_{t,xt}(\wn,\qn)&=\frac{\qn^{3}\wn}{(\wn^{2} - \qn^{2})^{2}} G_{\mbox{\tiny{mix}}}(\wn,\qn)\, ,& G_{x,xt}(\wn,\qn)&=-\frac{\qn^{2}\wn^{2}}{(\wn^{2} - \qn^{2})^{2}}G_{\mbox{\tiny{mix}}}(\wn,\qn) \, .\label{Gxxt}
\end{align}

\noindent Naturally, the matrix of Green's functions is symmetric, namely $G_{xx,tt}(\wn,\qn)=G_{tt,xx}(\wn,\qn)$, $G_{t,tt}(\wn,\qn)=G_{tt,t}(\wn,\qn)$ and so forth. The form of the correlators above is consistent with the field theory Ward identities (see \cite{Herzog:2009xv,Kovtun:2005ev}, for example), and it agrees with previously found expressions in the simpler case where there is no mixing between the electromagnetic and gravitational perturbations \cite{Miranda:2008vb}. It is worth noticing that the apparent pole at $\wn = \pm \qn$ in the expressions for the correlators is spurious, and reflects only the choice of normalization of $G_{1}$ and $G_{2}$. 
%%%%%%%%%%%%%%%%%%%%%%%%%%%%%%%%%%%%%%%%%%%%%%%%%%%%%%%%%%%%%%%%%%%%%%%%%%%%%%%%%%%%%%%%%%
\subsection{Small frequency expansion}\label{small frequency expansion}
In this subsection we obtain analytical results for the small frequency behavior of the retarded Green's functions, by expanding $\hat \Psi_{\pm}(\wn,\qn)$ and $\hat \Pi_{\pm}(\wn,\qn)$ for small $\wn$. We write $\Psi_{\pm}(u)$ as a power series in $\wn$,
\begin{align}\label{outerPhi}
\Psi_{O\pm}(u)&=\Psi^{(0)}_{O\pm}(u)+\wn \Psi^{(1)}_{O\pm}(u)+\wn^2 \Psi^{(2)}_{O\pm}(u)+\cdots\, ,
\end{align}

\noindent with the subscript denoting the outer region of the background geometry (see the discussion below). At extremality, extra care is needed in taking the limit $\wn \to 0$ near the horizon. This is due to the fact that in the extremal case $f(u)$ has a double pole at the horizon. A convenient way of handling the limit was introduced in \cite{Faulkner:2009wj}. Basically, one realizes that near the horizon the equations \eqref{sphipmeq} organize themselves as functions of $\zeta\equiv \omega\eta$, where, using \eqref{definition eta} and \eqref{dimensionless},  
\beq\label{inner}
u=1+\frac{\wn}{\sqrt{12}\zeta}\, .
\eeq

\noindent Since the coordinate $\zeta$ is the suitable radial coordinate in the $\ads_2$ factor of the near horizon region, the $\wn$ expansion of  $\Psi_{\pm}$ in that domain is given by 
\begin{align}\label{innerPhi}
\Psi_{I\pm}(\zeta)&=\Psi^{(0)}_{I\pm}(\zeta)+\wn \Psi^{(1)}_{I\pm}(\zeta)+\wn^2 \Psi^{(2)}_{I\pm}(\zeta)+\cdots\, , 
\end{align}

\noindent with the subscript $I$ denoting that the fields $\Psi_{\pm}$ are expanded as a power series in $\wn$ in the near horizon geometry (inner region). After imposing infalling boundary conditions for $\Psi_{I\pm}$ at the horizon, we match the inner and outer expansions in the so-called ``matching region" where the  $\zeta\to 0$ and $\wn/\zeta\to 0$ limits are taken. Since equations \eqref{sphipmeq} are linear, if we require that the solutions for the higher order terms in the expansions \eqref{outerPhi} and \eqref{innerPhi} do not include terms proportional to  the zeroth-order solutions near the matching region, we just need to match $\Psi^{(0)}_{I\pm}(\zeta)$ to $\Psi^{(0)}_{O\pm}(u)$ \cite{Faulkner:2009wj}. 
%%%%%%%%%%%%%%%%%%%%%%%%%%%%%%%%%%%%%%%%%%%%%%%%%%%%%%%%%%%%%%%%%%%%%%%%%%%%%%%%%%%%%%%%%%
\subsubsection{Inner region}
Near the horizon, we find that $\Psi^{(0)}_{I\pm}(\zeta)$ satisfy    
\begin{align}\label{leadinglinner}
-\Psi^{(0)\prime\prime}_{I\pm}(\zeta)+\left(\frac{\qn^2+2\mp2\sqrt{1+\qn^2}}{2\zeta}-1\right)\Psi^{(0)}_{I\pm}(\zeta)=0\, ,
\end{align}

\noindent which are identical to the equations of motion one obtains for two massive scalar fields in AdS$_2$, with masses $m_{\pm}$ given by
\begin{align}\label{msfeq}
m_\pm ^2 L_2^2&=1+\frac{\qn^2}{2}\mp\sqrt{1+\qn^2}\, .
\end{align}

\noindent Note that  $L_2$ in \eqref{msfeq} is  the curvature radius of AdS$_2$. 

Assuming that there exists a CFT dual to the AdS$_2$ region, one can think of $\Psi^{(0)}_{I\pm}(\zeta)$ as being dual to scalar operators ${\cal O}_{\pm}$ in the IR CFT. According to the standard AdS/CFT dictionary,  the  conformal dimensions $\delta_{\pm}=\nu_{\pm}+\frac{1}{2}$ of the operators ${\cal O}_{\pm}$ are related to the masses $m_{\pm}$ of the scalar fields $\Psi^{(0)}_{I\pm}$ by 
\begin{align}\label{nupm}
\nu_{\pm}=\frac{1}{2}\sqrt{1+4m_\pm ^2 L_2^2}=\frac{1}{2}\sqrt{5+2\qn^2\mp 4\sqrt{1+\qn^2}}\, .
\end{align}

\noindent Next, we impose infalling boundary conditions for $\Psi^{(0)}_{I\pm}(\zeta)$ at $\zeta\to\infty$. Having done so, near the matching region (where one takes the $\zeta\to 0$ and $\wn/\zeta\to 0$ limits) $\Psi^{(0)}_{I\pm}(\zeta)$ take the form  
\begin{align}\label{inzerothphipm}
\Psi^{(0)}_{I\pm}(u\to1)=\left[(u-1)^{-\frac{1}{2}+\nu_{\pm}}+{\cal G}_{\pm}(\wn)(u-1)^{-\frac{1}{2}-\nu_{\pm}}\right],
\end{align}

\noindent where we used  \eqref{inner} to change $\zeta$ back to $u$. Notice that a specific normalization for $\Psi^{(0)}_{I\pm}$ has been assumed in writing \eqref{inzerothphipm}. Clearly, such a choice does not affect the calculation of the boundary theory Green's functions. The functions ${\cal G}_{\pm}(\wn)$ in \eqref{inzerothphipm}  denote the retarded Green's functions of the IR CFT operators ${\cal O}_{\pm}$, and are given by \cite{Faulkner:2009wj, Edalati:2009bi}
\begin{align}\label{tworetgreen}
{\cal G}_\pm(\wn)=-2\nu_\pm e^{-i\pi\nu_\pm}\frac{\G(1-\nu_\pm)}{\G(1+\nu_\pm)}\left(\frac{\wn}{2}\right)^{2\nu_\pm}\, .
\end{align}
%%%%%%%%%%%%%%%%%%%%%%%%%%%%%%%%%%%%%%%%%%%%%%%%%%%%%%%%%%%%%%%%%%%%%%%%%%%%%%%%%%%%%%%%%%
\subsubsection{Outer region and matching}
In the outer region, the equations for $\Psi^{(0)}_{O\pm}(u)$ are obtained by setting $\wn=0$ in \eqref{sphipmeq}. The solutions for $\Psi^{(0)}_{O\pm}(u)$ are given by a linear combination of $(u-1)^{-\frac{1}{2}+\nu_{\pm}}$ and $(u-1)^{-\frac{1}{2}-\nu_{\pm}}$ near the matching region. Let us define
\begin{align}
\eta^{(0)}_{\pm}(u)=(u-1)^{-\frac{1}{2}+\nu_{\pm}}+\cdots\, , \qquad \qquad \xi^{(0)}_{\pm}(u)=(u-1)^{-\frac{1}{2}-\nu_{\pm}}+\cdots\, , \qquad u\to 1\, .
\end{align}

\noindent Then, matching  $\Psi^{(0)}_{O\pm}(u)$ to \eqref{inzerothphipm} we obtain
\begin{align}
\Psi^{(0)}_{O\pm}(u)=\left[\eta^{(0)}_{\pm}(u)+{\cal G}_\pm(\wn)\xi^{(0)}_{\pm}(u)\right].
\end{align}

Following the discussions in \cite{Faulkner:2009wj, Edalati:2010hk}, to higher orders we can write
\begin{align}
\eta_{\pm}(u)&=\eta^{(0)}_{\pm}(u)+\wn \eta^{(1)}_{\pm}(u)+\wn^2 \eta^{(2)}_{\pm}(u)+\cdots\, ,\\
\xi_{\pm}(u)&=\xi^{(0)}_{\pm}(u)+\wn \xi^{(1)}_{\pm}(u)+\wn^2 \xi^{(2)}_{\pm}(u)+\cdots\, , 
\end{align}

\noindent where $\eta^{(n>0)}_{\pm}(u)$ and $\xi^{(n>0)}_{\pm}(u)$ are obtained demanding that in the $u\to 1$ limit, they are distinct from $\eta^{(0)}_{\pm}(u)$ and $\xi^{(0)}_{\pm}(u)$, respectively. Thus, 
\begin{align}
\Psi_{O\pm}(u)=\left[\eta_{\pm}(u)+{\cal G}_\pm(\wn)\xi_{\pm}(u)\right].
\end{align}

\noindent Near $u\to \infty$, one can expand $\eta^{(n)}_{\pm}(u)$ and $\xi^{(n)}_{\pm}(u)$ as follows (here $n\geq 0$)
\begin{align}
\eta^{(n)}_{\pm}(u\to\infty)&=a^{(n)}_{\pm} \Big(1+\cdots\Big)+b^{(n)}_{\pm} \frac{1}{u}\Big(1+\cdots\Big)\, ,\\
\xi^{(n)}_{\pm}(u\to\infty)&=c^{(n)}_{\pm} \Big(1+\cdots\Big)+d^{(n)}_{\pm}\frac{1}{u}\Big(1+\cdots\Big)\, ,
\end{align}

\noindent where the coefficients $a^{(n)}_\pm, b^{(n)}_\pm, c^{(n)}_\pm$ and  $d^{(n)}_\pm$  are all functions of $\qn$. Thus, asymptotically we obtain
\begin{align}\label{asymPhipm}
\Psi_{O\pm}(u\to \infty)\simeq \hat\Psi_{\pm}\left(1+\frac{\hat\Pi_{\pm}}{u}\right),
\end{align}

\noindent where
\begin{align}\label{xpm}
\hat\Psi_{\pm}&=\left[a^{(0)}_\pm+\wn a^{(1)}_\pm+{\cal O}\left(\wn^2\right)\right]+ {\cal G}_\pm(\wn)\left[ c^{(0)}_\pm+\wn c^{(1)}_\pm+{\cal O}\left(\wn^2\right)\right],\\
\hat\Psi_{\pm}\hat\Pi_{\pm} &=\left[b^{(0)}_\pm+\wn b^{(1)}_\pm+{\cal O}\left(\wn^2\right)\right]+ {\cal G}_\pm(\wn)\left[ d^{(0)}_\pm+\wn d^{(1)}_\pm+{\cal O}\left(\wn^2\right)\right].\label{ypm}
\end{align}
%%%%%%%%%%%%%%%%%%%%%%%%%%%%%%%%%%%%%%%%%%%%%%%%%%%%%%%%%%%%%%%%%%%%%%%%%%%%%%%%%%%%%%%%%%
\subsection{Criticality: emergent IR scaling}
In general, the coefficients $a^{(n)}_\pm, b^{(n)}_\pm, c^{(n)}_\pm, d^{(n)}_\pm$  should be computed numerically. Once these coefficients are known, one can substitute the expressions for $\hat\Psi_{\pm}$ and $\hat\Psi_{\pm}\hat\Pi_{\pm}$ into \eqref{Gtttt}--\eqref{Gxxt} in order to compute the retarded Green's functions. In the small frequency (but generic momentum) regime, the Green's functions can be analyzed semi-analytically even though the coefficients $a^{(n)}_\pm, b^{(n)}_\pm, c^{(n)}_\pm, d^{(n)}_\pm$ are not known explicitly. Since the coefficients $a^{(n)}_\pm, b^{(n)}_\pm, c^{(n)}_\pm$, $d^{(n)}_\pm$ are all real, the complex parts of $\hat\Psi_{\pm}$ and $\hat\Psi_{\pm}\hat\Pi_{\pm}$ are determined by the IR CFT Green's functions ${\cal G}_\pm (\wn)$. Assuming that the product $a^{(0)}_{+}a^{(0)}_{-}$ does not vanish for generic values of momentum $\qn$, expanding the denominator $\hat\Psi_{+}\hat\Psi_{-}$ in \eqref{G1} and \eqref{G2}, and noticing that  $2\nu_{-}>3$ while $2\nu_{+}>1$, we obtain the following small frequency scaling behavior for  ${\rm Im} ~G_1(\wn,\qn)$ and ${\rm Im} ~G_2(\wn,\qn)$: 
\begin{align}\label{sfImG1}
{\rm Im} ~G_1(\wn,\qn)&=\qn^{2}\left(1+\sqrt{1+\qn^2}\right)e_{0}(\qn)~{\rm Im}~ {\cal G}_{+}(\wn)\left[1+\cdots\right]\propto \wn^{2\nu_{+}},\\
{\rm Im} ~G_2(\wn,\qn)&=\frac{8}{\mu^{2}}\left(1-\sqrt{1+\qn^2}\right)e_{0}(\qn)~{\rm Im}~ {\cal G}_{+}(\wn)\left[1+\cdots\right]\propto \wn^{2\nu_{+}},\label{sfImG2}
\end{align}

\noindent where the dots represent  subleading terms which vanish as $\wn\to 0$, and 
\begin{align}
e_{0}(\qn)=\frac{9C}{2}\frac{\qn^{2}}{\sqrt{1+\qn^2}}\frac{b^{(0)}_{+}}{a^{(0)}_{+}}\left(\frac{d^{(0)}_{+}}{b^{(0)}_{+}}-\frac{c^{(0)}_{+}}{a^{(0)}_{+}}\right).
\end{align}

\noindent As a result, the matrix of the spectral functions of the boundary theory operators $\hat J_t$, $\hat J_x$,  $\hat T_{tt}$,  $\hat T_{xx}$, $\hat T_{yy}$,  $\hat T_{xt}$  shows scaling behavior at small frequency
\begin{equation}
\begin{array}{lll}
{\rm Im} ~G_{tt,tt}\propto \wn^{2\nu_{+}},&\qquad  {\rm Im} ~G_{xx,tt}\propto \wn^{2+2\nu_{+}},&\qquad {\rm Im} ~G_{yy,tt}\propto \wn^{2\nu_{+}},\\
{\rm Im} ~G_{xt,tt}\propto \wn^{1+2\nu_{+}},&\qquad {\rm Im} ~G_{xx,xx}\propto \wn^{4+2\nu_{-}},&\qquad {\rm Im} ~G_{yy,xx}\propto \wn^{2+2\nu_{+}},\\
{\rm Im} ~G_{xt,xx}\propto \wn^{3+2\nu_{+}},&\qquad {\rm Im} ~G_{yy,yy}\propto \wn^{2\nu_{+}},&\qquad {\rm Im} ~G_{xt,yy}\propto \wn^{1+2\nu_{+}},\\
{\rm Im} ~G_{xt,xt}\propto \wn^{2+2\nu_{+}},&\qquad {\rm Im} ~G_{t,tt}\propto \wn^{2\nu_{+}},&\qquad {\rm Im} ~G_{x,tt}\propto \wn^{1+2\nu_{+}},\\
{\rm Im} ~G_{t,xx}\propto \wn^{2+2\nu_{+}},&\qquad {\rm Im} ~G_{x,xx}\propto \wn^{3+2\nu_{+}},&\qquad {\rm Im} ~G_{t,yy}\propto \wn^{2\nu_{+}},\\
{\rm Im} ~G_{x,yy}\propto \wn^{1+2\nu_{+}},&\qquad {\rm Im} ~G_{t,xt}\propto \wn^{1+2\nu_{+}},&\qquad {\rm Im} ~G_{x,xt}\propto \wn^{2+2\nu_{+}},\\
{\rm Im} ~G_{t,t}\propto \wn^{2\nu_{+}},&\qquad {\rm Im} ~G_{x,t}\propto \wn^{1+2\nu_{+}},&\qquad {\rm Im} ~G_{x,x}\propto \wn^{2+2\nu_{+}}.
\end{array} 
\end{equation}

Notice that the low-frequency scaling behavior of the above spectral functions is emergent: the scaling is a consequence of the near horizon geometry (which translates into the IR physics of the boundary theory) containing an AdS$_2$ factor. Although $e_0$ depends on details of the background geometry in UV, the scaling behavior $\wn^{2\nu_{+}}$ does not, as long as the AdS$_2$ part of the geometry is unchanged.
%%%%%%%%%%%%%%%%%%%%%%%%%%%%%%%%%%%%%%%%%%%%%%%%%%%%%%%%%%%%%%%%%%%%%%%%%%%%%%%%%%%%%%%%%%
%%%%%%%%%%%%%%%%%%%%%%%%%%%%%%%%%%%%%%%%%%%%%%%%%%%%%%%%%%%%%%%%%%%%%%%%%%%%%%%%%%%%%%%%%%
\section{The spectrum of the boundary theory}\label{section: spectrum}
As we have discussed, the set of poles of the Green's functions \eqref{Gtttt}--\eqref{Gxxt} in the complex frequency plane, hereafter denoted by $\wn^{\pm}_{\star}(\qn)$, are generically  given by the solution of $\hat\Psi_{\pm}(\wn,\qn)=0$ (with infalling boundary conditions at the horizon). Thus, we have set up the problem in such a way that finding the spectrum of the boundary theory amounts to finding the quasinormal frequencies of $\Psi_{\pm}$ in the RN-AdS$_{4}$ black hole  background: in the context of the AdS/CFT correspondence, this should be regarded as the definition of the quasinormal frequencies in asymptotically AdS geometries \cite{Horowitz:1999jd, Birmingham:2001pj, Son:2002sd, Kovtun:2005ev}, thus removing the ambiguity associated with the choice of asymptotic boundary conditions which is inherent to the eigenvalue problem in the gravity side.

Aside from its connection to the dual field theory retarded Green's functions, finding the electromagnetic and gravitational  quasinormal frequencies of the extremal RN-AdS$_{4}$ black hole is an interesting problem in its own right. From the gravity point of view, knowledge of the QNM spectra allows one to construct the dissipative response of the black hole geometry to small perturbations. In the case of the non-extremal RN-AdS$_{4}$ black hole, such quasinormal frequencies were numerically computed in \cite{Berti:2003ud}. For the extremal RN-AdS$_{4}$ black hole, the  quasinormal frequencies of the odd-parity (electromagnetic and gravitational) modes  (which are dual to the operators in the shear channel of the boundary theory) were recently computed  in \cite{Edalati:2010hk}. In this section,  we extend our previous computation to the case where the perturbations are even under the $y\to-y$ parity reflection. Using gauge/gravity duality, these fluctuations are mapped to operators in the sound channel of the dual field theory. See \cite{Berti:2009kk} for an extensive review of the subject of quasinormal modes in general relativity and the AdS/CFT correspondence. 
%%%%%%%%%%%%%%%%%%%%%%%%%%%%%%%%%%%%%%%%%%%%%%%%%%%%%%%%%%%%%%%%%%%%%%%%%%%%%%%%%%%%%%%%%%
\subsection{QNMs: matrix method}
Although there exist various semi-analytic methods for calculating small or large (magnitudes of) quasinormal frequencies, it is often necessary to perform numeric calculations in order to extract their generic values. Following \cite{Edalati:2010hk,Denef:2009yy}, we use a discrete algorithm developed by \cite{Leaver:1990zz} in order to compute the sound-channel quasinormal frequencies of the coupled electromagnetic and gravitational perturbations of the extremal RN-AdS$_{4}$ black hole. For numerical purposes it is convenient to switch to a new radial coordinate $z=1/u$. In terms of this coordinate the black hole horizon is located at $z=1$, while the asymptotic boundary is located at $z=0$.  The equations \eqref{equation for Psi} then read
\begin{equation}\label{eq for Psi in terms of z}
z^{2}\left[f(z)\Psi^{\prime}_\pm(z)\right]'-z^{2}J_\pm(z)\Psi^{\prime}_\pm(z)+K_\pm(z)\Psi_\pm(z)=0 \, ,
\end{equation}

\noindent where the prime now denotes derivatives with respect to $z$, and $f(z)$ is given by
\begin{equation}
f(z) = 1 - (1 + Q^{2})z^{3} + Q^{2}z^{4}\, .
\end{equation}

\noindent Similarly, $K_{\pm}(z)$ and $J_{\pm}(z)$ can be read off from \eqref{definition J and K}-\eqref{definition J and K 2}. In the method of \cite{Leaver:1990zz}, one first factorizes the leading behaviors of the normalizable modes of $\Psi_{\pm}(z)$ at the horizon and the boundary (which in practice amounts to imposing asymptotic Dirichlet boundary conditions). What remains is then approximated by a power series in $z$ around a point $z_0$, chosen in such a way that the radius of convergence of the series reaches both the horizon and the boundary. In our case we can take $z_0=\frac{1}{2}$. Thus, we write
\begin{align}\label{seriesPhipm}
\Psi_{\pm}(z)=e^{i\frac{\wn}{\sqrt{12}(1-z)}}f(z)^{-i\frac{\sqrt{3}\wn}{9}} z \sum_{m=0}^M a^{\pm}_m(\wn,\qn) \left(z-\frac{1}{2}\right)^m\,.
\end{align}

\noindent The first two prefactors are chosen to select the infalling wave at the extremal black hole horizon, as it can be easily seen by rewriting them in terms of infalling Eddington coordinates. Substituting \eqref{seriesPhipm} into \eqref{eq for Psi in terms of z}, we obtain a set of $M+1$ linear equations for $M+1$ unknowns $\{a_p(\wn,\qn)\}$, that can be written in matrix form as 
\begin{align}\label{definition matrix A}
\sum_{m=0}^M A^{\pm}_{mp}(\wn,\qn) a^{\pm}_p(\wn,\qn)=0\, .
\end{align}

Suppose now we choose a specific value for the spatial momentum $\qn$. The quasinormal frequencies $\wn^{\pm}_{\star}$  for that particular value of $\qn$ are then solutions of
\begin{align}\label{det}
\det~A^{\pm}(\wn^{\pm}_\star,\qn)=0\,.
\end{align}

\noindent Since the fields $\Psi_{\pm}$ in \eqref{seriesPhipm}  are approximated by a power series, the more terms kept in the series, the more accurate the results obtained for the quasinormal frequencies. The quasinormal spectra for $\qn = 0.5$ obtained as described above is shown in Figure \ref{Figq05} below. 

\begin{figure}[h]
\begin{center}
$\begin{array}{cc}
~~~~~~~~(a)&~~~~~~~~~~~~~~(b)\\ \\
\includegraphics[width=2.9in]{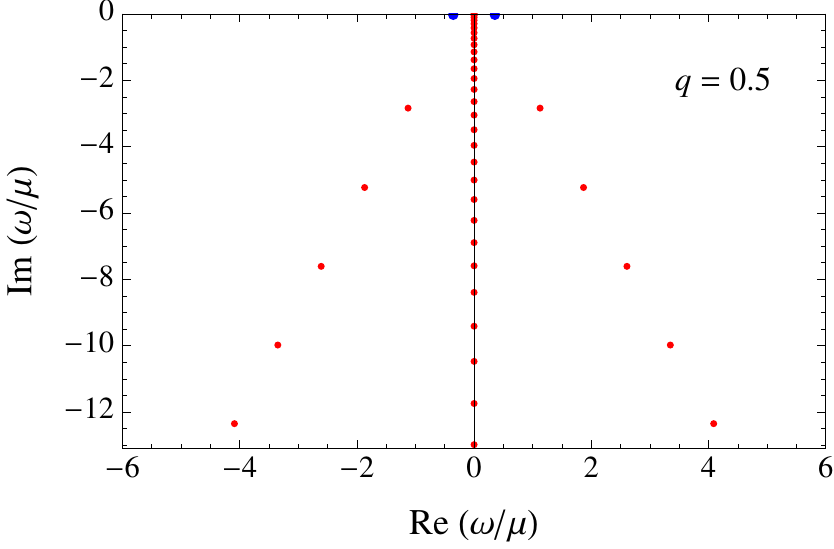}&\qquad
\includegraphics[width=2.9in]{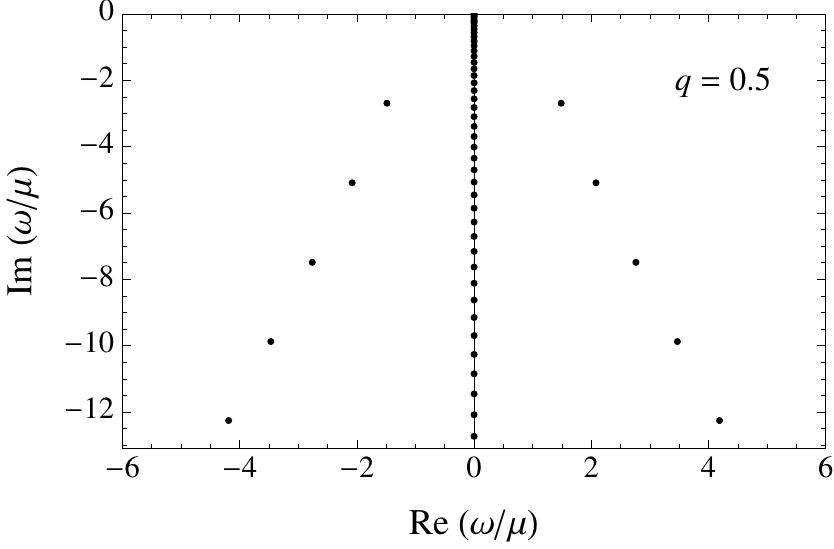}
\end{array}$\\
$\quad\quad~~~(c)$\\ 
$\phantom{a}$\\
\includegraphics[width=3.5in]{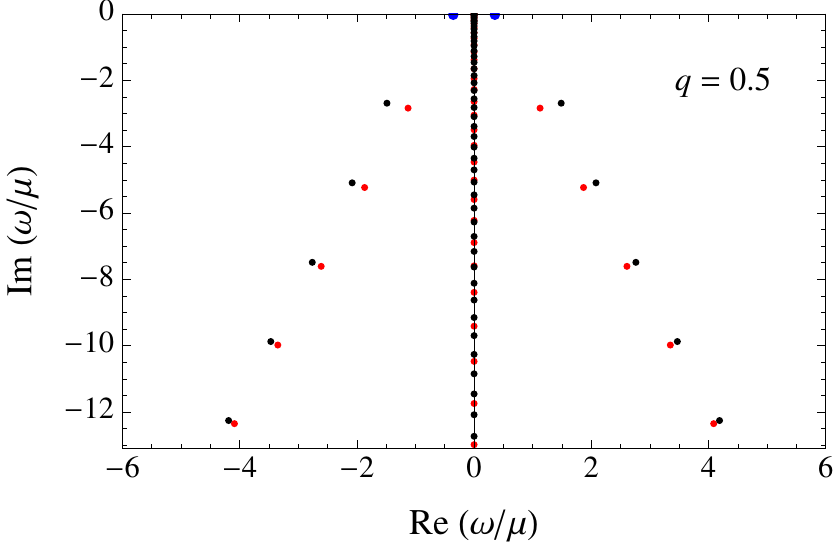}
\end{center}
\caption[FIG. \arabic{figure}.]{Sound-channel electromagnetic and gravitational quasinormal frequencies for the extremal Reissner-Nordstr\"{o}m  AdS$_4$ black hole.  Plot (a) shows the quasinormal frequencies of $\Psi_{+}$, with the sound modes depicted in blue. Plot (b) shows the quasinormal frequencies of $\Psi_{-}$. For both plots, $\qn=0.5$ and $M=300$. Plot (c) shows the spectra of $\Psi_{+}$ and $\Psi_{-}$ superimposed.}
   \label{Figq05}
\end{figure}
A distinctive feature of the spectrum at zero temperature is that the series of purely imaginary frequencies on the negative imaginary axis, which are evenly spaced at finite temperature,\footnote{This has been determined by an explicit calculation of the quasinormal frequencies at finite temperature, which for the sake of brevity we do not display here.} now coalesce to form a branch cut. Indeed, this was observed in the correlators of conserved currents in the shear channel \cite{Edalati:2010hk} as well as in those of chargeless scalar operators \cite{Denef:2009yy}. As discussed in \cite{Edalati:2010hk}, at zero spatial momentum ($\qn = 0$) this feature is explained by the explicit appearance of logarithmic terms of the form $\wn^{n} \log \wn$ (with $n\in \mathds{Z}$) in the Green's functions, while for $\qn \neq 0$ it is due to terms of the form $\wn^{2\nu_{+}}$ (c.f. \eqref{sfImG1} and \eqref{sfImG2}), where $2\nu_{+}$ is generically an irrational number. On the gravity side, one can understand the appearance of the branch cut by using the tortoise coordinate to write the equations for $\Psi_{\pm}$ in a Schr\"odinger-like form. Then, the non-analyticity is due to the fact that the effective potential near the horizon decays as a power law rather than exponentially (as would be the case at finite $T$), a typical feature of extremal black holes (see \cite{Ching:1995tj} and references therein). That we can still observe a discrete series of poles as we move down the imaginary axis is an artifact of the numerical algorithm. As we increase the size $M$ of the matrix $A$ defined in \eqref{definition matrix A}, these poles move upwards and become dense over a larger set of imaginary frequencies. The plots show only the lower half of the complex-$\wn$ plane, because no instabilities exist. This was also the case for the shear-channel modes, where moreover it was relatively easy to prove the stability analytically, at least over a broad range of momenta \cite{Edalati:2010hk}. 

Besides the coalescent poles giving rise to the branch cut, we observe a discrete series of quasinormal frequencies with a non-vanishing real part. Notice that they appear in conjugate pairs with the same imaginary part, but with opposite sign for the real part. This was to be expected due to the unbroken parity symmetry of the dual theory, a feature that would be absent in the presence of a magnetic field, for example. In figure \ref{Figq05} (a) we can distinctively appreciate the lowest lying pair (depicted in blue). As we will explicitly confirm below by analyzing their dispersion relation, this pair of poles correspond to the sound modes (in the sense described in section \ref{zero temperature hydro}: they satisfy $\omega(k\to 0) = \pm c_{s}k - i\Gamma_{s}k^{2}$ and dominate the decay of the longitudinal fluctuations over long wavelengths and time scales). While the imaginary part of these quasinormal frequencies is non-zero, it cannot be appreciated in the figures above as a consequence of the rather small attenuation constant $\Gamma_{s}$ we will find below. The rest of the QNMs with a non-vanishing real part are the so-called \textit{overtones}, whose dispersion relation will seen to be quite different from that of the sound modes.  They correspond to higher resonances whose effects we expect to become noticeable as we go beyond the small frequency and momentum approximation. As it can be observed in figure \ref{Figq05} (c), the overtones of $\Psi_{+}$ and $\Psi_{-}$ coalesce as we move down in the complex frequency plane. 
%This feature can be understood as follows: as we go to large (compared to $\mu$) values of the frequency the effects of the charge density fade away, which %means that the spectrum effectively approaches that of the chargeless black hole. Since there is only one master field in the chargeless case (c.f. appendix %\ref{chargeless}), we expect the two series of overtones to come together into the overtones of the single master field whose effects remain important at large %energies/low charge density.
%%%%%%%%%%%%%%%%%%%%%%%%%%%%%%%%%%%%%%%%%%%%%%%%%%%%%%%%%%%%%%%%%%%%%%%%%%%%%%%%%%%%%%%%%%
\subsection{Dispersion relations: speed of sound and sound attenuation}
\begin{figure}[h!]
$\begin{array}{cc}
~~~~~(a)&~~~~~~~~~~~~~~(b)\\ \\
\includegraphics[width=2.86in]{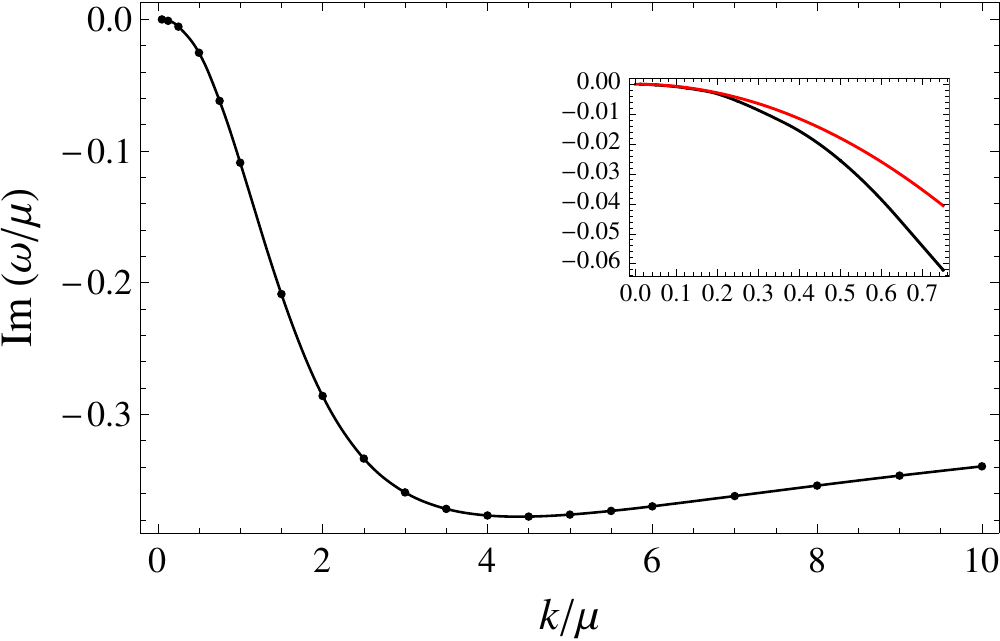}&\qquad
\includegraphics[width=2.86in]{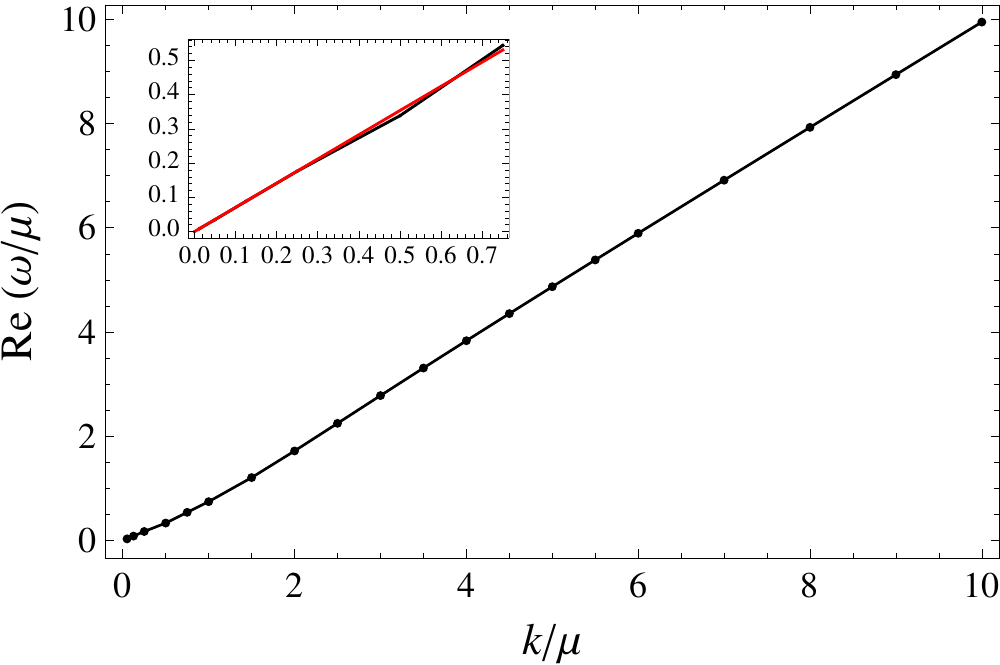}
\end{array}$
\caption[FIG. \arabic{figure}.]{\footnotesize{Dispersion relation of the sound modes. Plot (a) depicts the imaginary part of $\wn_{s}$ as a function of momenta, while plot (b) shows the real part of $\wn_{s}$. The dots represent the actual values obtained numerically, while the solid line results from fitting the data by a smooth curve. The inset plots compare the expected dispersion relation \eqref{expected sound dispersion} (red line) with the values obtained numerically (black line), for small values of the momenta. For large momentum, the slope of the real part becomes unity within numerical precision.}}
   \label{ExtSound}
\end{figure}

At finite temperature, we can easily compute the speed of sound $c_s$ in the boundary field theory by using the one-point function of the energy-momentum tensor operator $\langle T_{\mu\nu} \rangle$. The only non-vanishing components of the second fundamental form in our background are
\begin{equation}\label{components extrinsic curvature}
\bar{K}^{t}_{\phantom{t}t} = -\frac{1}{2}\frac{(2f + uf')}{L\sqrt{f}}, \qquad \bar{K}^{x}_{\phantom{x}x} = \bar{K}^{y}_{\phantom{y}y}=-\frac{\sqrt{f}}{L}\, ,
\end{equation}

\noindent and with this result we can evaluate \eqref{one point function stress tensor} to obtain 
\begin{align}
\langle T^{t}_{\phantom{t}t}\rangle &\equiv - \epsilon = -\frac{r_{0}^{3}}{\kappa_{4}^{2}L^{4}}M\, ,\\
\langle T^{x}_{\phantom{x}x}\rangle &=\langle T^{y}_{\phantom{y}y}\rangle   \equiv P = \frac{\epsilon}{2} = \frac{r_{0}^{3}}{2\kappa_{4}^{2}L^{4}}M\, ,
\end{align}

\noindent where  $\epsilon$ is the energy density that had been introduced in \eqref{ece} and $P = \epsilon/2$ is the pressure. Notice that the expectation value of the trace of the stress-energy tensor $\langle T^{i}_{\phantom{i}i}\rangle$ is zero. Substituting the values of $\epsilon$ and $P$ into the thermodynamic relation $c_s^2=\partial P/\partial \epsilon$ we obtain
\begin{equation}
c_s^2=\frac{\partial P}{\partial \epsilon} = \frac{1}{2}\, ,
\end{equation}

\noindent which is the expected result in two spatial dimensions. Note, however, that this result is independent of whether the RN-AdS$_{4}$ black hole background is extremal or non-extremal. Thus, for our zero-temperature dual field theory at finite U(1) charge density, the speed of sound should also be $1/\sqrt{2}$. On the other hand, calculating the QNM spectrum for various values of $\qn$ we can numerically construct the dispersion relation of the sound modes. The result is shown in figure \ref{ExtSound}.
 \noindent We can observe that both the real and imaginary parts of $\wn_{s}$ approach zero as $\qn \to 0$.  Moreover, for small values of the momenta the dispersion relation has the form \eqref{sound modes dispersion}; fitting the numeric data for $\qn \leq 0.25$, we obtain $c_{s} \approx 0.704$ and $\mu \Gamma_{s} \approx 0.083$. Notice the agreement between the numerically obtained value for the speed of sound and the value $c_{s}=1/\sqrt{2} \approx 0.707$ predicted by (the zero-temperature limit of) hydrodynamics. To gain some insight about the value obtained for the sound attenuation constant $\Gamma_{s}$, we can again investigate the $T\to 0$ limit of well-known hydrodynamical relations. As discussed in section \ref{zero temperature hydro}, such a procedure may or may not yield the right value for the theory at zero temperature. Putting aside this issue for the moment, using the ratio $s/\epsilon =2\pi Q/(\mu M)$ (see \eqref{ece}) we find
\begin{equation}\label{attenuation at finite temp}
\mu\, \Gamma_{s} = \frac{\mu}{2}\frac{\eta}{\epsilon + P} = \frac{1}{3}\frac{\eta}{s}\frac{\mu s}{\epsilon} = \frac{2\pi}{3}\frac{\eta}{s}\frac{Q}{M}\, .
\end{equation}

\noindent At extremality, $Q^{2}=3$, $M = 4$, and the ratio of the shear viscosity $\eta$ to entropy density $s$ was found to be $\eta/s = 1/(4\pi)$ \cite{Edalati:2009bi}, just as in the well-known finite temperature examples (see \cite{Son:2007vk} and references therein). Thus, by taking the $T\to 0$ limit of \eqref{attenuation at finite temp} we expect
\begin{equation}\label{expected sound attenuation}
\left. \mu \,\Gamma_{s}\right|_{T\to 0} = \frac{1}{8\sqrt{3}} \approx 0.072\, . 
\end{equation}

\noindent Summarizing, by taking the zero temperature limit of hydrodynamics we anticipate the sound modes (the one with positive real part, say) to have a dispersion relation of the form 
\begin{equation}\label{expected sound dispersion}
\wn_{s}(\qn \to 0) = \frac{\qn}{\sqrt{2}} -i\frac{\qn^{2}}{8\sqrt{3}}\, .
\end{equation}
\noindent The inset plots in figure \ref{ExtSound} compare the dispersion relation \eqref{expected sound dispersion} (red line) with the one obtained from the actual numerical data at $T=0$ (black line). While the agreement in the real part of the frequency prevails in a relatively broad range of momenta, the imaginary part computed by the different methods agrees only in the low frequency and momentum regime.  In particular, the actual value $\mu \Gamma_{s} \approx 0.083$ for the sound attenuation constant differs slightly from the expected value \eqref{expected sound attenuation}. This discrepancy might be attributed to numerical uncertainties, but we note that it is rather large, of order 13\%, in fact much larger than the uncertainty implied by the results for the speed of sound. 

\begin{figure}[h!]
$\begin{array}{cc}
~~~~~(a)&~~~~~~~~~~~~~~(b)\\ \\
\includegraphics[width=2.86in]{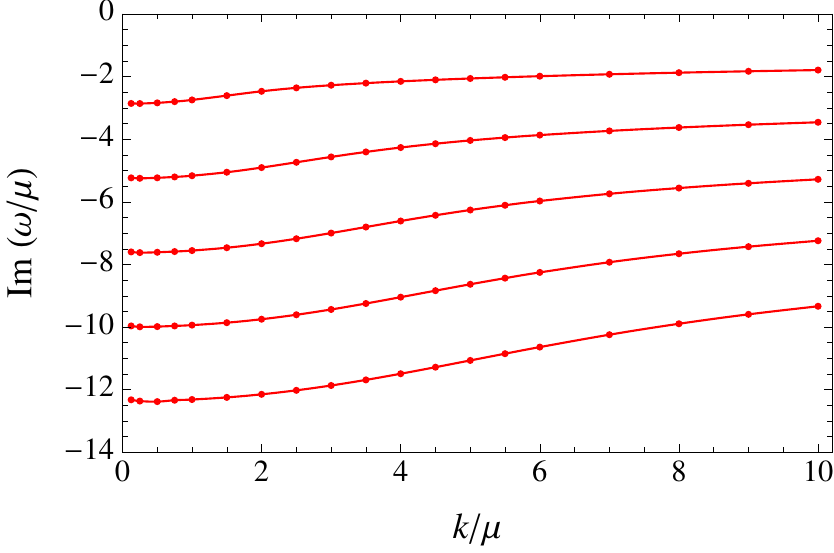}&\qquad
\includegraphics[width=2.86in]{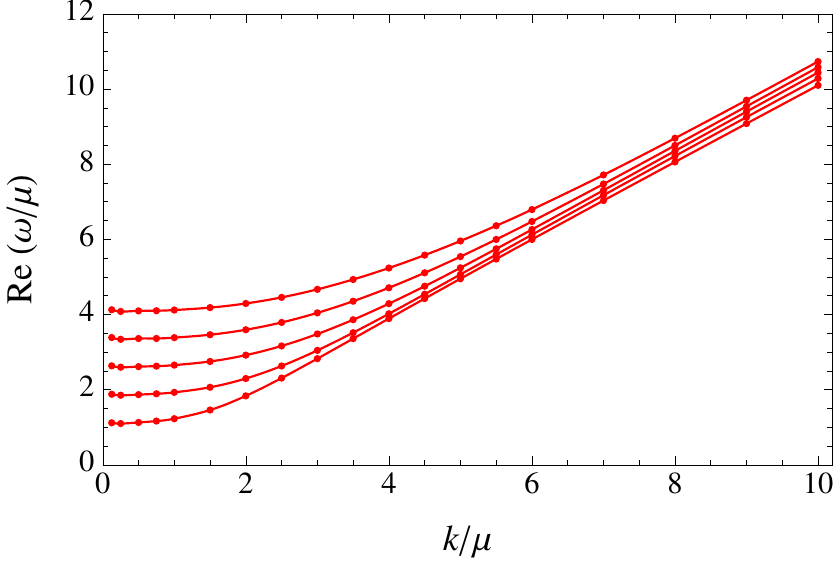}
\end{array}$
\caption[FIG. \arabic{figure}.]{\footnotesize{Dispersion relation of the first five overtones of $\Psi_{+}$. Plot (a) shows the variation of the imaginary part of the corresponding quasinormal frequencies as we vary the momentum $\qn$. Plot (b) shows the corresponding change in the real part of the frequency.}}
   \label{OvertonesPlus}
\end{figure}

Turning now to the overtones, we can also follow their trajectories in the complex frequency plane as we vary the momentum $\qn$ and determine in this way their dispersion relation. Figure \ref{OvertonesPlus} shows the result for the first five overtones in the spectrum of $\Psi_{+}$. 

\begin{figure}[h!]
$\begin{array}{cc}
~~~~~(a)&~~~~~~~~~~~~~~(b)\\ \\
\includegraphics[width=2.86in]{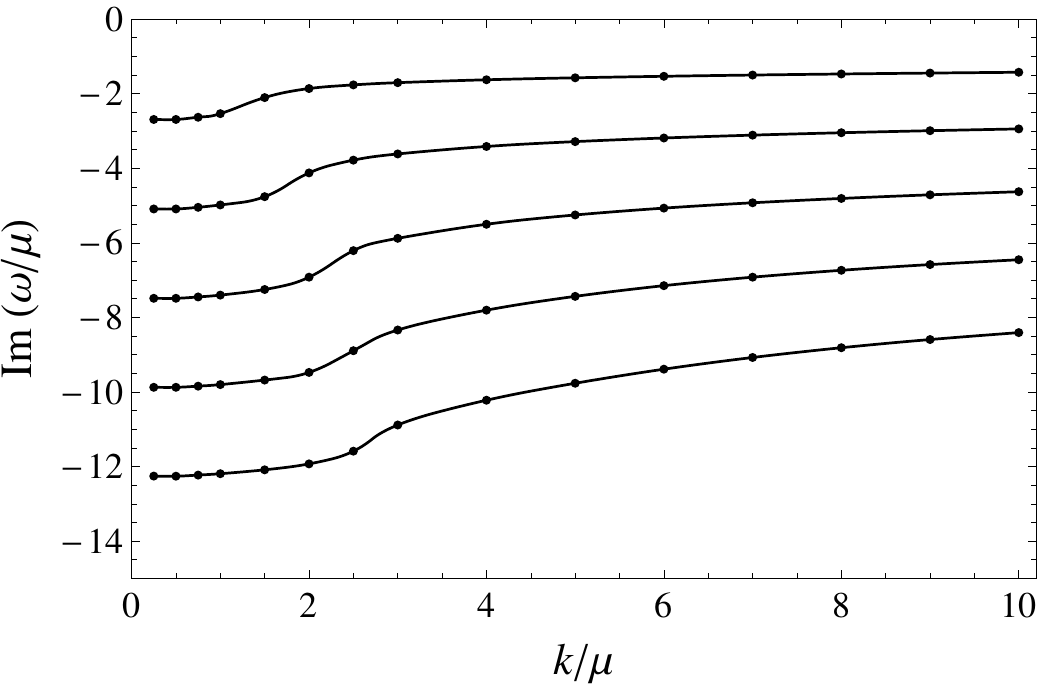}&\qquad
\includegraphics[width=2.86in]{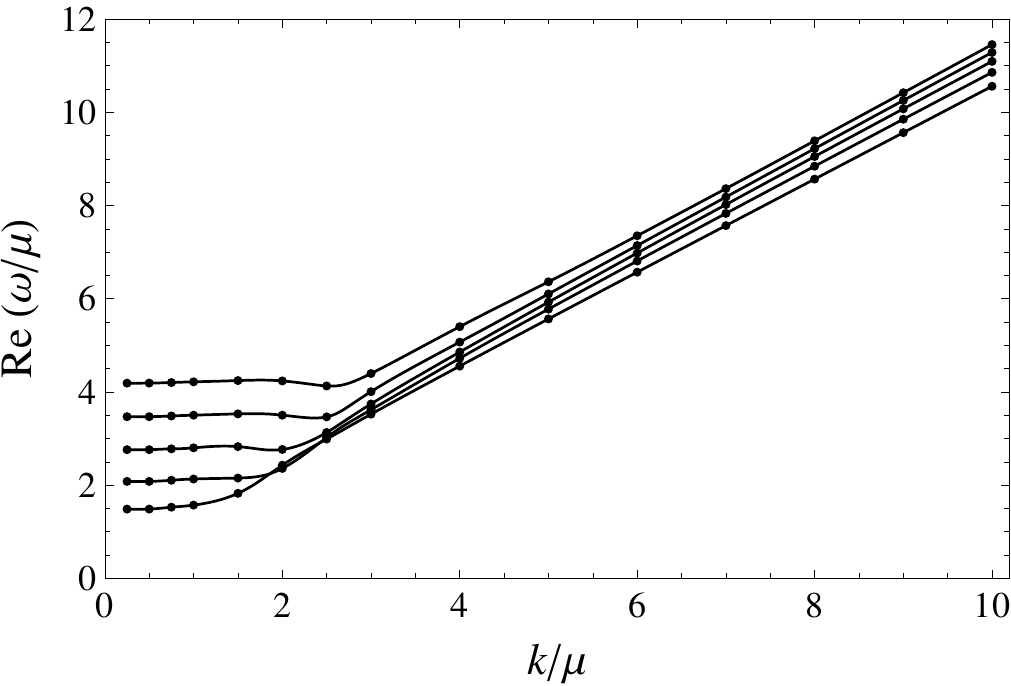}
\end{array}$
\caption[FIG. \arabic{figure}.]{\footnotesize{Dispersion relation of the first five overtones of $\Psi_{-}$. Plot (a) shows the variation of the imaginary part of the corresponding quasinormal frequencies as we vary the momentum $\qn$. Plot (b) shows the corresponding change in the real part of the frequency.}}
   \label{OvertonesMinus}
\end{figure}

\noindent As mentioned above, these frequencies do not approach zero as $\qn \to 0$, so these modes are not ``hydrodynamic" in the broad sense. In particular, notice that the imaginary part of the quasinormal frequencies is approximately constant for small values of the momenta with $\qn < 1$ ($k < \mu$), and also for large values of the momenta of the order $\qn \sim 10$. In contrast, the real part of these overtones is a monotonically increasing function of $\qn$, satisfying a linear dispersion with unit slope (within the numerical precision) for values of the momenta greater than $\qn \sim 7$. This feature is explained by the fact that at large $\qn$ and $\wn$ the influence of the charge density becomes less important, and the background effectively approaches a Lorentz-invariant vacuum. Provided that the higher resonances behave like sharp, well-defined particle-like excitations at large momenta, their dispersion relation becomes  effectively relativistic in the UV. The same qualitative features are observed for the first five overtones of $\Psi_{-}$, whose dispersion relations are shown in figure \ref{OvertonesMinus}.

%%%%%%%%%%%%%%%%%%%%%%%%%%%%%%%%%%%%%%%%%%%%%%%%%%%%%%%%%%%%%%%%%%%%%%%%%%%%%%%%%%%%%%%%%%
\subsection{Residues}
In order to further understand the field theory spectrum, it is important to compute the residues of the retarded correlators at the various quasinormal frequencies, as they determine the weight with which they contribute to the spectral function. To compute the residues numerically, we found it convenient to use a method combining series expansions and numeric integration. Basically, the idea is to approximate  by a power series the solution for the fields $\Psi_{\pm}(z)$ near the horizon (i.e. for $z \geq 1-\epsilon$) and near the boundary (i.e. for $z \leq \delta$), and integrate numerically from $z=\delta$ to $z = 1-\epsilon$. In this way, one avoids the stiffness issues associated with numeric integration near the singularities.\footnote{This method was used in \cite{Friess:2006kw} to compute the quasinormal spectrum. In particular, the quasinormal frequencies are obtained as the zeros of the Wronskian between the numeric solution and the series approximation at the matching point.} In order to ensure convergence of the series expansions, for the values of the momenta we have studied it was sufficient to choose $\epsilon = 10^{-3}$ and $\delta = 10^{-7}$, and we kept ten terms in the series approximations at each end of the integration interval. By matching the numerically obtained solution for the fields and their derivatives with the series expansions at $z =\delta$, we can construct the coefficients $\Psi^{(0)}_{\pm} \equiv \hat{\Psi}_{\pm}$ and $\Psi^{(1)}_\pm$ in \eqref{asymptotics of Psipm} as a function of the frequency (for fixed $\qn$) and with them we obtain the canonical momenta $\Pi_{\pm}$ using \eqref{momenta in terms of source}. At this step we can compute the residue of $\Pi_{\pm}$ at a given quasinormal frequency $\wn_{\star}$, which is given by
\begin{equation}
\mbox{Res }\Pi_{\pm}(\wn_{\star})= \left.\frac{1}{C_{\pm}}\frac{\Psi^{(1)}_{\pm}(\wn)}{ \partial_{\wn}\hat{\Psi}_{\pm}(\wn)}\right|_{\wn = \wn_{\star}}\, .
\end{equation}

\noindent In practice, the denominator in this expression can be evaluated by discretizing the $\wn$-derivative. From the residue of $\Pi_{\pm}$ we can calculate $\mbox{Res }G_{1}$ and $\mbox{Res }G_{2}$ using \eqref{G1}-\eqref{G2}, which in turn allows us to compute the residue of the actual retarded Green's functions constructed from $G_{1}$ and $G_{2}$ as given in \eqref{Gtttt}--\eqref{Gxxt}. 

\begin{figure}[h!]
\begin{center}
$\begin{array}{cc}
~~~~~(a)&~~~~~~~~~~~~(b)\\ \\
\includegraphics[width=2.7in]{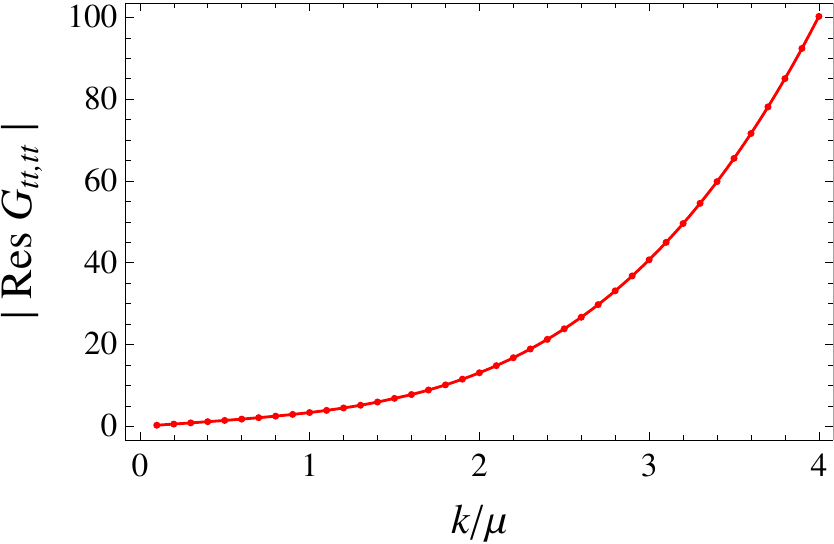}&\qquad
\includegraphics[width=2.7in]{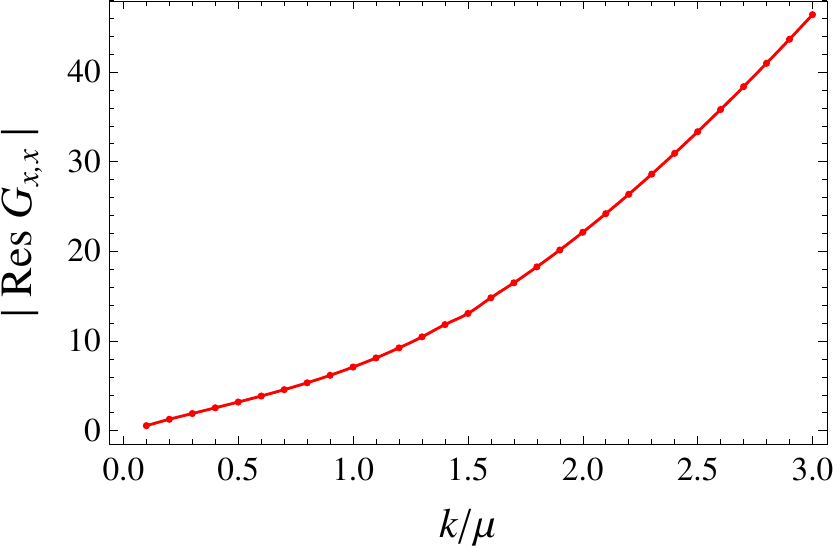}
\end{array}$
\end{center}
\caption[FIG. \arabic{figure}.]{\footnotesize{Absolute value of the sound pole residue as a function of momentum. Plot (a) shows the residue of $G_{tt,tt}$ ; plot (b) shows the residue of $G_{x,x}$. }}
   \label{ResiduesttttAndxx}
\end{figure}

In principle one should not only compute the residues of the sound modes, but those of the higher resonances as well. For example, knowledge of the residue at the first overtone allows one to compute the relaxation time scale into the hydrodynamic regime \cite{Amado:2007yr,Amado:2008ji}. While being cognizant of the importance of these residues, due to the complexity of the numerical calculations involved, here we will restrict ourselves to the hydrodynamic (sound) modes. As a representative of our results, in figure \ref{ResiduesttttAndxx} we display the absolute value of the sound pole residue for $G_{tt,tt}$ and $G_{x,x}$. 

\noindent  In particular, we observe that the absolute value of the residue approaches zero as $\qn \to 0$, and behaves as a monotonically increasing function of $\qn$.
%%%%%%%%%%%%%%%%%%%%%%%%%%%%%%%%%%%%%%%%%%%%%%%%%%%%%%%%%%%%%%%%%%%%%%%%%%%%%%%%%%%%%%%%%%
\subsection{Green's functions}
By using the same combination of series approximations and numeric integration that was employed in the calculation of the residues, one can numerically construct the retarded Green's functions \eqref{Gtttt}-\eqref{Gxxt} as a function of frequency (for fixed $\qn$). Since we have also computed the sound pole residue, one can compare the Green's functions obtained numerically with an approximation of the form \eqref{Breit Wigner in sound channel}. The results for the imaginary and real parts of $G_{tt,tt}$ at $\qn = 0.5$ and $\qn = 1$ are shown in figure \ref{Gtttt q 05 and 1}. It is worth mentioning that the contact terms presented in appendix \ref{appendix: contact} should be taken into account when plotting the real part of the retarded correlators. 

\begin{figure}[h!]
\begin{center}
$\begin{array}{cc}
~~~~~(a)&~~~~~~~~~~~~(b)\\ \\
\includegraphics[width=2.95in]{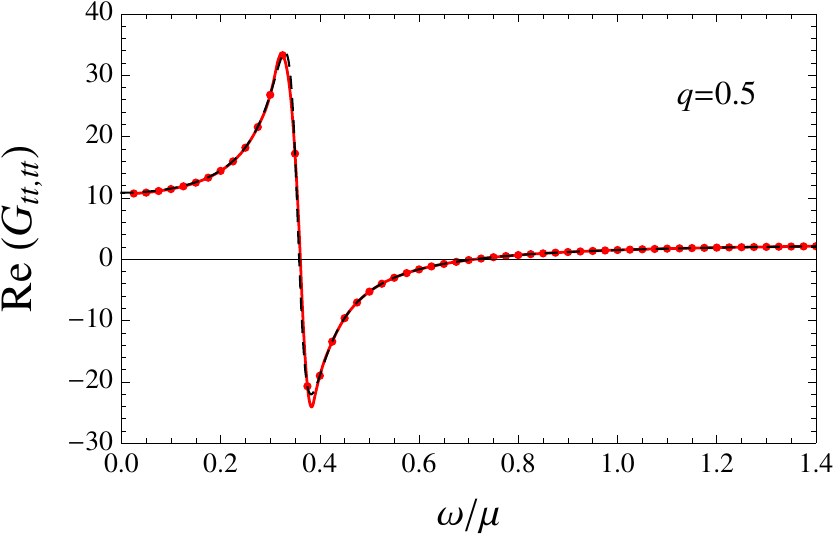}&\qquad
\includegraphics[width=2.95in]{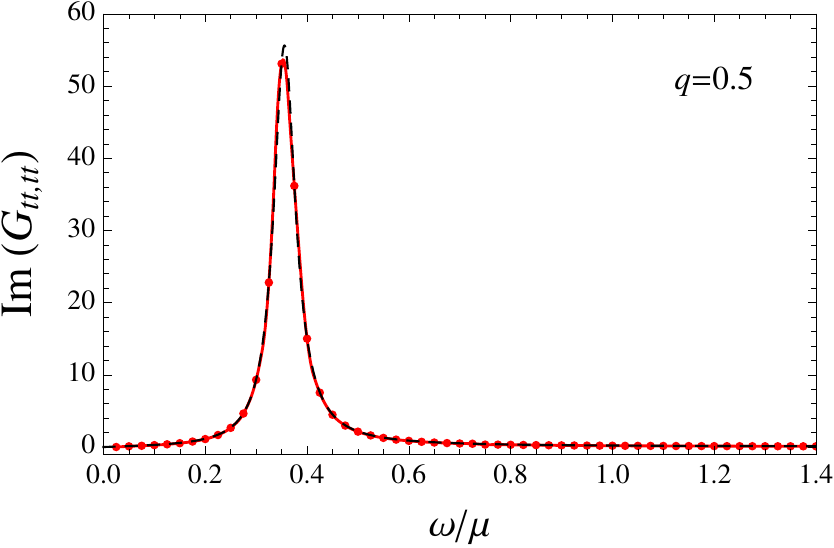}\\ \\
~~~~~(c)&~~~~~~~~~~~~(d)\\ \\
\includegraphics[width=2.95in]{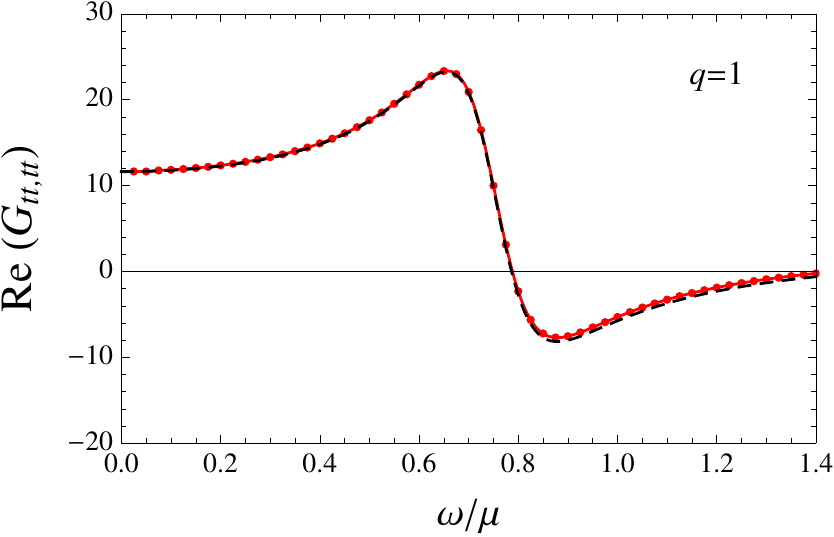}&\qquad
\includegraphics[width=2.95in]{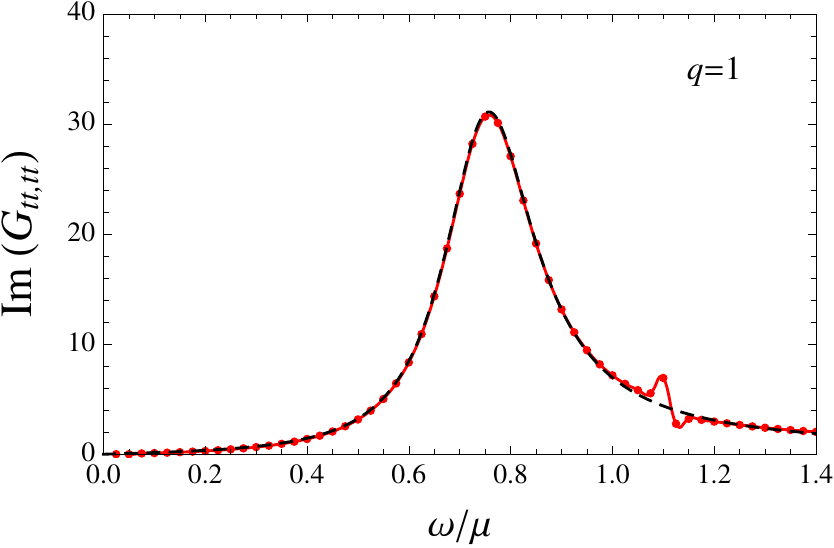}
\end{array}$
\end{center}
\caption[FIG. \arabic{figure}.]{\footnotesize{Real and imaginary parts of the $G_{tt,tt}$ Green's functions as a function of the (real part of the) frequency. Plots (a) and (b) correspond to $\qn = 0.5$, while plots (c) and (d) correspond to $\qn$ = 1. The dots represent the actual numerical data, and the solid (red) is the interpolation of the data by a smooth curve. The dashed (black) line represents the Green's function approximated by \eqref{Breit Wigner in sound channel}, with the spectrum truncated to include the sound modes only.}}
   \label{Gtttt q 05 and 1}
\end{figure}
At $\qn = 0.5$ (plots (a) and (b)) we observe an excellent agreement between the Green's function obtained numerically and the approximation obtained by truncating the spectrum to the sound modes only. Moreover, the sound pole resonance is sharply peaked, so that these modes are well-defined particle-like excitations. This is natural, since at small values of $\qn$ the modes with $\omega(k\to 0)\to 0$ are expected to dominate the spectrum. For $\qn =1$, the real part of the Green's function is also accurately described by the sound mode. In contrast, the numerically computed imaginary part shows two smaller peaks between $\wn = 1.0$ and $\wn = 1.2$, which are not captured by the truncated approximation including the sound mode only (dashed line). As we increase the momentum, we should in general expect the contribution of the higher resonances to become noticeable, and it is therefore tempting to attribute these peaks to the effects of the first overtones of $\Psi_{+}$ and $\Psi_{-}$, respectively. However, the position of the peaks in the plot slightly differs from the first overtone frequencies as computed using Leaver's method. This might indicate that the numeric precision in our computations might not be adequate to  conclusively identify the effects of higher resonances. A possible way of elucidate the nature of these peaks is to include the contribution of the first overtones in the spectral approximation \eqref{Breit Wigner in sound channel} and compare it with the curve obtained from the numerics. Naturally, this approach presumes a knowledge of the residues of said poles with high enough accuracy, which we were not able to obtain with the numeric techniques described above. 
%%%%%%%%%%%%%%%%%%%%%%%%%%%%%%%%%%%%%%%%%%%%%%%%%%%%%%%%%%%%%%%%%%%%%%%%%%%%%%%%%%%%%%%%%%
%%%%%%%%%%%%%%%%%%%%%%%%%%%%%%%%%%%%%%%%%%%%%%%%%%%%%%%%%%%%%%%%%%%%%%%%%%%%%%%%%%%%%%%%%%
\section{Conclusions}\label{section: conclusions}
Using gauge/gravity duality techniques, we have constructed the correlation functions of the conserved charge current and energy-momentum tensor of a (2+1)-dimensional strongly coupled field theory at finite density and zero temperature. This problem is more challenging  than the cases addressed previously in the literature: not only we have considered the coupled electromagnetic and gravitational fluctuations, but we also faced the additional complications introduced by the more severe singularity structure of the equations of motion in the extremal background. In the process of these numerical computations, we resolved a number of technical obstacles, and in particular were able to set up the Dirichlet problem for decoupled master fields that determine the correlation functions of interest. We have found that the correlation functions at finite density and zero temperature are essentially hydrodynamic in character at low frequency and momentum, with the chemical potential setting the scale (rather than temperature). In the IR, they satisfy scaling behavior originating from the IR CFT dual to the near-horizon geometry, and at generic momenta display a dominant sound mode and a sequence of resonances (with typically small residue), along with a branch cut at $\omega=0$. It would be interesting to apply these techniques to other systems of direct physical interest.

\vskip 2cm
\centerline{\bf Acknowledgments}

It is a pleasure to thank S. Baharian, E. Fradkin, B. Goodman, J. McGreevy, L. Pando-Zayas, J.C. Peng and P. Phillips for helpful conversations and correspondence. R.G.L.\ and M.E.\ are supported by DOE grant FG02-91-ER40709, J.I.J. is supported by a Fulbright-CONICYT fellowship.
%%%%%%%%%%%%%%%%%%%%%%%%%%%%%%%%%%%%%%%%%%%%%%%%%%%%%%%%%%%%%%%%%%%%%%%%%%%%%%%%%%%%%%%%%%
%%%%%%%%%%%%%%%%%%%%%%%%%%%%%%%%%%%%%%%%%%%%%%%%%%%%%%%%%%%%%%%%%%%%%%%%%%%%%%%%%%%%%%%%%%
\appendix
%%%%%%%%%%%%%%%%%%%%%%%%%%%%%%%%%%%%%%%%%%%%%%%%%%%%%%%%%%%%%%%%%%%%%%%%%%%%%%%%%%%%%%%%%%
%%%%%%%%%%%%%%%%%%%%%%%%%%%%%%%%%%%%%%%%%%%%%%%%%%%%%%%%%%%%%%%%%%%%%%%%%%%%%%%%%%%%%%%%%%
\section{Master field equations}\label{appendix: master fields}
In this section we provide further details about the master fields introduced in the body of the paper.
%%%%%%%%%%%%%%%%%%%%%%%%%%%%%%%%%%%%%%%%%%%%%%%%%%%%%%%%%%%%%%%%%%%%%%%%%%%%%%%%%%%%%%%%%%
\subsection{Charged black hole}
The Einstein-Maxwell equations \eqref{eqt}--\eqref{equx} can be reduced to two decoupled second-order differential equations for the so-called master fields $\Phi_\pm(u)$ \cite{Kodama:2003kk}. In order to obtain the master fields, define first
\begin{align}
{\cal A}(u)&=u^2a'_t(u)+\mu{h^y}_y(u)-\frac{1}{2}\mu{h^t}_t(u)\,,\\
\Phi(u)&=u{h^y}_y(u)-H(u)^{-1}u^4 f(u)\left[{h^x}_x'(u)+{h^y}_y'(u)\right]\, ,\label{Phiu}
\end{align}

\noindent where $H(u)=u^3f'(u)+Q^2\qn^2$. Equations \eqref{eqt}--\eqref{equx} now reduce to a set of two coupled second-order differential equations
\begin{align}
&\left[u^2f(u){\cal A}'(u)\right]'+2\mu f(u)\Phi'(u)+C_1(u) {\cal A}(u)+\mu C_2(u) \Phi(u)=0,\label{eqA}\\
&\left[u^2f(u)\Phi'(u)\right]'+C_3(u) \Phi(u)+\mu^{-1}Q^2C_4(u) {\cal A}(u)=0\label{eqPhi},
\end{align}

\noindent with
\begin{align}
C_1(u)&=\frac{Q^2}{u^2f(u)}\left(\wn^2-f(u)\qn^2-\frac{8f(u)^2}{H(u)}\right),\\
C_2(u)&=\frac{1}{H(u)}\left(\left[Q^2\qn^2-10u^2f(u)\right]f'(u)+24uf(u)[1-f(u)]+\frac{Q^4\qn^4}{u^3}\right),\\
C_3(u)&=\frac{1}{u^2f(u)}\Big(Q^2\wn^2-f(u)V(u)\Big),\\
C_4(u)&=-\frac{4}{H(u)^2}\left(u^2f'(u)\left[10f(u)+uf'(u)\right]+24uf(u)\left[f(u)-1\right]-\frac{Q^4\qn^4}{u^3}\right),
\end{align}

\noindent where 
\begin{align}\label{defV}
V(u)&=\frac{1}{u^2H(u)^2}\Big\{9u^{11}f'(u)^3+u^6\left[3(22f(u)-8)u^4+Q^2\left(9\qn^2u^2+4\right)\right]f'(u)^2\nonumber\\
&\phantom{=}+u^3\left[216u^6f(u)(f(u)-1)+8Q^2\qn^2u^4(2f(u)-3)+Q^4\qn^2\left(\qn^2u^2+8\right)\right]f'(u)\nonumber\\
&\phantom{=}+288 u^8f(u)(f(u)-1)^2-24Q^2\qn^2u^6f(u)(f(u)-1)+Q^6\qn^4\left(\qn^2u^2+4\right)\Big\}.
\end{align}

\noindent Equations \eqref{eqA} and \eqref{eqPhi} can then be decoupled  by introducing the master fields 
\begin{align}\label{Phipmddef}
\Phi_\pm(u)=\alpha_{\pm}(u) \Phi(u) +\mu^{-1}Q^2{\cal A}(u)\, ,
\end{align}

\noindent where
\begin{align}
\alpha_{\pm}(u)= \frac{1}{2}F_\pm(Q)-\frac{Q^2}{u}, \qquad F_\pm(Q)=\frac{3}{4}\Big[(1+Q^2)\pm\sqrt{(1+Q^2)^2+(16/9)Q^4\qn^2}\Big].
\end{align}

\noindent To obtain the equations for the master fields, solve \eqref{Phipmddef} for ${\cal A}(u)$ and $\Phi(u)$ in terms of $\Phi_\pm(u)$ to obtain
\begin{align}
\mu^{-1}Q^2{\cal A}(u)&=\gamma\left[\alpha_{+}(u)\Phi_{-}(u)-\alpha_{-}(u)\Phi_{+}(u)\right],\label{APhipm}\\
\Phi(u)&=\gamma\left[\Phi_{+}(u)-\Phi_{-}(u)\right],\label{PhiPhipm}
\end{align}

\noindent where
\begin{align}
\gamma\equiv\frac{1}{\alpha_{+}(u)-\alpha_{-}(u)}=\frac{4}{3\sqrt{(1+Q^2)^2+(16/9)Q^4\qn^2}}\, .
\end{align}

\noindent Substituting \eqref{APhipm} and \eqref{PhiPhipm} into \eqref{eqA} and \eqref{eqPhi}, we arrive at
\begin{align}
&\left[u^2f(u)\Phi'_\pm(u)\right]' \pm\gamma\left[Q^2f'(u)+\alpha_{\pm}(u)C_1(u)-Q^2C_2(u)-\alpha_{\pm}C_3(u)+\alpha^2_{\pm}(u)C_4(u)\right]\Phi_{\mp}(u)\nonumber\\
&\mp \gamma\left[Q^2f'(u)+\alpha_{\mp}(u)C_1(u)-Q^2C_2(u)-\alpha_{\pm}C_3(u)+\alpha_{+}(u)\alpha_{-}(u)C_4(u)\right]\Phi_{\pm}(u) =0\, .
\end{align}

\noindent Note that
\begin{align}\label{cond}
Q^2f'(u)+\alpha_{\pm}(u)C_1(u)-Q^2C_2(u)-\alpha_{\pm}(u)C_3(u)+\alpha^2_{\pm}(u)C_4(u)=0\, .
\end{align}

\noindent Simplifying the equations for the master fields, we arrive at
\begin{align}\label{gPhipmeq}
u^2f(u)\left[u^2f(u)\Phi'_\pm(u)\right]'+ \left[Q^2\wn^2-U_{\pm}(u)\right]\Phi_{\pm}(u)=0\, ,
\end{align}

\noindent where 
\begin{align}
U_{\pm}(u)&=\pm \gamma f(u)\left\{\alpha_{\pm}(u)V(u)+\frac{Q^2}{4\alpha_{\pm}(u)}\left[Q^2\qn^2H(u)+8Q^2f(u)\right]+\frac{2Q^2}{u}\left[H(u)-2Q^2\qn^2\right]\right. \nonumber\\ 
&~~~~~~~~~~~~~~~~\left. +\frac{Q^2}{H(u)}\left[48u^3f(u)(f(u)-1)+20u^4f(u)f'(u)\right]\right\}.
\end{align}
%%%%%%%%%%%%%%%%%%%%%%%%%%%%%%%%%%%%%%%%%%%%%%%%%%%%%%%%%%%%%%%%%%%%%%%%%%%%%%%%%%%%%%%%%%
\subsection{Chargeless case} \label{chargeless}
In the case where the background in chargeless, \ie $Q=0$, the gravitational and electromagnetic perturbations decouple, and the fields $\Phi_{\pm}$ reduce to a single master field. In our notation, this field is given by $\Phi_{+}(u)=\Phi(u)$ and it encodes the information from the gravity sector only. It can be easily seen to satisfy
\begin{align}\label{ucmastereq}
u^2f(u)\left[u^2f(u)\Phi'(u)\right]'+ \left[9w^2-U(u)\right]\Phi(u)=0\, , 
\end{align}

\noindent where 
\begin{align}
U(u)= \frac{f(u)}{u\left(1+3q^2u\right)^2}\left[2u^3f(u)+3\left(1+3q^2u\right)\left(1+9q^4u^2\right)\right],
\end{align}

\noindent and
\begin{align}
w\equiv \frac{\omega}{4\pi T}\, ,\qquad \quad q\equiv\frac{k}{4\pi T}\, .
\end{align}
%%%%%%%%%%%%%%%%%%%%%%%%%%%%%%%%%%%%%%%%%%%%%%%%%%%%%%%%%%%%%%%%%%%%%%%%%%%%%%%%%%%%%%%%%%
%%%%%%%%%%%%%%%%%%%%%%%%%%%%%%%%%%%%%%%%%%%%%%%%%%%%%%%%%%%%%%%%%%%%%%%%%%%%%%%%%%%%%%%%%%
\section{Contact terms}\label{appendix: contact}
Here we display explicitly the contact terms that supplement the Green's functions \eqref{Gtttt}-\eqref{Gxxt}. For notational simplicity, in the expressions below we suppress the overall constant $C$ coming from the normalization of the boundary action. We display the results for arbitrary temperature (equivalently, $Q$), but it is understood that in the calculations of the paper we have used the extremal value $Q =\sqrt{3}$. The contact term contributions to the $\langle T_{\mu\nu}T_{\rho\sigma}\rangle$ correlators are\footnote{The name ``contact term" usually denotes contributions which are analytic in frequency and momentum. As noticed in the body of the paper, the apparent pole at $\wn = \pm \qn$ is spurious and denotes a choice of normalization. In this sense, the expressions presented in this appendix are to be regarded as true (analytic) contact terms. }
\begin{align}
C_{tt,tt}&= -\frac{(1+Q^{2})}{2}\frac{2\wn^{4} - 7\qn^{2}\wn^{2}+ 8 \qn^{4}}{(\wn^{2} - \qn^{2})^{2}}\, ,& C_{xx,tt}&=\frac{(1+Q^{2})}{2}\frac{\wn^{4} - 2\qn^{2}\wn^{2}-2 \qn^{4}}{(\wn^{2} - \qn^{2})^{2}}\, ,\\
C_{yy,tt}&=\frac{(1+Q^{2})}{2}\frac{\wn^{2} +2 \qn^{2}}{\wn^{2} - \qn^{2}}\, ,& C_{xt,tt}&=-\frac{3(1+Q^{2})}{2}\frac{\qn\wn(\wn^{2} -2 \qn^{2})}{(\wn^{2} - \qn^{2})^{2}}\, ,\\
C_{xx,xx}&=\frac{(1+Q^{2})}{2}\frac{\wn^{4} - 5\qn^{2}\wn^{2}+  \qn^{4}}{(\wn^{2} - \qn^{2})^{2}}\, ,&  C_{yy,xx}&=\frac{(1+Q^{2})}{2}\frac{2\wn^{2} +\qn^{2}}{\wn^{2} - \qn^{2}}\, ,\\ 
C_{xt,xx}&=\frac{3(1+Q^{2})}{2}\frac{\qn^{3}\wn}{(\wn^{2} - \qn^{2})^{2}}\, ,& C_{yy,yy}&= \frac{(1+Q^{2})}{2}\, ,\\
C_{xt,yy}&= -\frac{3(1+Q^{2})}{2}\frac{\qn\wn}{\wn^{2} - \qn^{2}}\, ,& C_{xt,xt}&=(1+Q^{2})\frac{2\wn^{4} - 4\qn^{2}\wn^{2}- \qn^{4}}{(\wn^{2} - \qn^{2})^{2}}\, .
\end{align}

\noindent On the other hand, there is no contact term contribution to the current-current correlators of the form $\langle J_{\mu}J_{\nu}\rangle$:
\begin{align}
C_{t,t}(\wn,\qn)&=0\, , & C_{x,t}(\wn,\qn)&=0\, ,& C_{x,x}(\wn,\qn)&=0\, .
\end{align}

\noindent For the ``mixed" correlators $\langle J_{\mu}T_{\nu\rho} \rangle$ we find
\begin{align}
C_{t,tt}&= - \frac{4Q^{2}}{\mu}\frac{\qn^{2}(\wn^{2} - 2\qn^{2})}{(\wn^{2} - \qn^{2})^{2}},& C_{x,tt}&= \frac{4Q^{2}}{\mu}\frac{\qn\wn(\wn^{2} - 2\qn^{2})}{(\wn^{2} - \qn^{2})^{2}},\\
C_{t,xx}&= \frac{4Q^{2}}{\mu}\frac{\qn^{4}}{(\wn^{2} - \qn^{2})^{2}}\, ,& C_{x,xx}&= -\frac{4Q^{2}}{\mu}\frac{\qn^{3}\wn}{(\wn^{2} - \qn^{2})^{2}}\, ,\\
C_{t,yy}&= - \frac{4Q^{2}}{\mu}\frac{\qn^{2}}{\wn^{2} - \qn^{2}}\, ,& C_{x,yy}&= \frac{4Q^{2}}{\mu}\frac{\qn\wn}{\wn^{2} - \qn^{2}}\, ,\\
C_{t,xt}&= \frac{4Q^{2}}{\mu}\frac{\qn\wn(\wn^{2} - 2\qn^{2})}{(\wn^{2} - \qn^{2})^{2}}\, ,& C_{x,xt}&= -\frac{4Q^{2}}{\mu}\frac{\wn^{2}(\wn^{2} - 2\qn^{2})}{(\wn^{2} - \qn^{2})^{2}}\, .
\end{align}

\newpage
%\bibliographystyle{JHEP}
%\bibliography{SoundBib}
\providecommand{\href}[2]{#2}\begingroup\raggedright\endgroup
\end{document}